\documentclass{JHEP3}
\usepackage{amsmath}
\usepackage{axodraw}
\usepackage{epsfig}
\usepackage{amssymb}
\usepackage{mathrsfs}
\usepackage{mathcomp}
\usepackage{cite}
\usepackage[latin1]{inputenc}

\def\rap{\ensuremath{Y}}

\keywords{QCD, Dipole model, Pomeron Loops, Saturation}
\preprint{LU-TP 06-35\\
  hep-ph/0610157\\
}

\title{Small-$x$ Dipole Evolution Beyond the Large-$N_c$ Limit}

\author{Emil Avsar, Gösta Gustafson and Leif Lönnblad\\
  Dept.~of Theoretical Physics,
  Sölvegatan 14A, S-223 62  Lund, Sweden\\
  E-mail: \email{Emil.Avsar@thep.lu.se}, \email{Gosta.Gustafson@thep.lu.se}
    and \email{Leif.Lonnblad@thep.lu.se}}
  
  \abstract{We present a method to include colour-suppressed effects
    in the Mueller dipole picture. The model consistently includes
    saturation effects both in the evolution of dipoles and in the
    interactions of dipoles with a target in a frame-independent way.
    
    When implemented in a Monte Carlo simulation together with our
    previous model of energy--momentum conservation and a simple
    dipole description of initial state protons and virtual photons,
    the model is able to reproduce to a satisfactory degree both the
    $\gamma^*p$ cross sections as measured at HERA as well as the
    total $pp$ cross section all the way from ISR energies to the
    Tevatron and beyond.}

\begin{document}

\sloppy

\section{Introduction}

Parton evolution at small $x$ is a difficult problem. It is interesting 
because of the strong nonlinear effects and the interplay between perturbative 
and non-perturbative physics, and it is an important problem, as it is 
necessary to have a good understanding of the dominant effects from
the strong interaction in the analyses of results from LHC and the 
interpretation of possible signals for new physics.

Although it cannot be possible to include all quantum interference effects 
in a classical branching process, such a stochastic evolution has been 
extremely successful to describe parton cascades in $e^+e^-$-annihilation. 
This is also the case in the DGLAP regime of DIS (high $Q^2$ and large $x$). 
In these cases the virtuality or transverse momentum acts as a well-defined evolution 
parameter, and the perturbative cascade can be well separated from the 
non-perturbative effects in the hadronization or the input distributions 
in the DGLAP evolution. We note, however, that important for the good 
agreement with LEP data is both a description of the hardest gluons by 
fixed order matrix elements, and implementation of energy--momentum 
conservation in the event simulation procedure.

DIS at small $x$ is more difficult. To LL or NLL accuracy the cross section 
is determined by the BFKL equation, which describes an evolution in $x$ 
instead of the $k_\perp$-ordering in the DGLAP evolution. A big problem is 
here that the NLL corrections are so large, that it in practice only can 
give a qualitative description of experimental data.

Part of the NLL corrections originate from the running coupling
$\alpha_s$.  Including a running coupling in the evolution implies
that the parton chain spreads into the non-perturbative region, and a
soft cutoff is needed for small $k_\perp$. This problem is avoided in
studies of collisions between highly virtual photons or massive onium
states. Another possibility is to study events with two high $p_\perp$
jets separated by a large rapidity interval. Also here it is
demonstrated that the energy--momentum conservation constraint has a
very large effect on the theoretical calculations
\cite{Andersen:2006sp}. In $ep$ and $pp$ scattering the influence of
non-perturbative effects cannot be avoided, however, and has to be
included in the analysis.

A most essential feature of the $e^+e^-$-cascades is the so-called
"soft colour interference" or angular ordering
\cite{Mueller:1981ex,Ermolaev:1981cm,Bassetto:1982ma,Marchesini:1983bm}.
A colour charge in one parton has always a corresponding anti-charge in
an accompanying parton, and it is important to account for the
interference in the radiation from these emitters. An efficient way to
treat this effect is offered by the dipole cascade model described in
refs.\ \cite{Gustafson:1986db,Gustafson:1988rq}, in which the QCD state
is described as a chain of colour dipoles formed by a
charge--anti-charge pair, rather than a chain of gluons. In the large
$N_c$ limit the dipoles radiate independently, and analyses of
experimental data from LEP \cite{Abbiendi:2003ri,Achard:2003ik,Schael:2006ns,
Siebel:2005uw}
indicate that the interference between
different dipoles is very weak. (In contrast to the other three LEP 
experiments, DELPHI \cite{Siebel:2005uw} does indeed favour some weak
effects from interaction.)

In DIS a formulation with coherent colour dipoles was presented by
Golec-Biernat--W\"usthoff\cite{Golec-Biernat:1998js,Golec-Biernat:1999qd}, and a space-like dipole cascade model was
formulated by Mueller
\cite{Mueller:1993rr,Mueller:1994jq,Mueller:1994gb}.  While the
time-like dipole cascade model in ref.~\cite{Gustafson:1988rq} is
formulated in momentum space, the dipoles in these models are
specified in transverse coordinate (or impact parameter) space. As the
transverse coordinate is little affected in a high energy interaction,
this makes it possible to account for multiple parton sub-collisions in
a natural way.

Mueller's dipole cascade 
model is valid in the large $N_c$ limit, and is demonstrated 
to satisfy the BFKL equation to LL accuracy. In this picture one starts with a 
$q\bar{q}$ colour singlet state (a quarkonium or simply onium state), and when 
the system is boosted to higher energy more and more soft gluons are emitted, 
forming a chain of colour dipoles. As mentioned this approach leads to the 
BFKL equation, but it also goes beyond the BFKL formalism. When two cascades 
collide it is possible to take into account multiple scattering to all orders, 
and it is thus possible to obtain a unitarised expression for the $S$-matrix. 
The probabilistic nature of the cascades implies that the evolution can be 
simulated in a Monte Carlo program, and the effects of unitarisation be studied 
numerically \cite{Salam:1996nb,Salam:1995uy} (see also \cite{Avsar:2004ms}).

The dipole model contains a vertex in which a dipole splits into two new 
dipoles, originating when a soft gluon is emitted from the original colour 
dipole. The evolution can then be formulated as a typical birth--death 
process, where a dipole can decay into two new dipoles with a specified 
differential probability, proportional to $dY$, with $Y=\log 1/x$
which here acts as a time variable for the evolution process.

There are some problems with the dipole evolution as formulated in 
\cite{Mueller:1993rr,Mueller:1994jq,Mueller:1994gb}.
One problem comes from the fact that the cascade cannot correspond to 
real gluon emissions. The splitting vertex diverges as the size of one of the 
new dipoles goes to zero. The many small dipoles interact, however, very 
weakly with a target, a phenomenon referred to as colour transparency. 
Thus, even if the 
number of small dipoles diverges, the total cross section remains finite. 
Although we thus get a finite cross section, the divergence causes problems: 
i) Dipoles which do not interact should be regarded as virtual. Therefore 
the dipole model in this formulation can be used to study fully inclusive 
quantities like the total cross section, but not the properties of exclusive 
final states. ii) In numerical calculations the divergence has to be regulated 
by a cutoff for small dipoles. Although the cutoff does not show up in the 
cross section, it makes it extremely time consuming to run a simulation 
program with a cutoff, which is small enough to simulate the physics with 
good accuracy.

As mentioned above, the constraint from energy--momentum 
conservation is very important in order to achieve agreement between 
theory and experiment in $e^+e^-$-annihilation, and in \cite{Andersen:2006sp} it is 
demonstrated that it also has a large effect in space-like cascades. 
This constraint goes beyond the LL approximation, 
and is not included in the formalism in \cite{Mueller:1993rr,Mueller:1994jq}. 
This is related to the 
problem with small dipoles discussed above. A very small dipole corresponds 
to well localized partons, which 
thus must have large transverse momenta, which in turn implies also 
non-negligible longitudinal momentum and energy. Only real final state 
partons have to obey conservation of energy and momentum, and to fully solve 
this problem one must also
solve the problem with specifying the final states.

Another problem is due to the fact that in this formalism dipoles in the 
same onium do not interact. Saturation effects are included in the collisions 
between two cascades, but not in the evolution of each cascade separately. 
This problem is related to the large $N_c$ approximation in the evolution. 
Multiple collisions are formally colour suppressed, and in the Lorentz frame 
where the collision is studied they lead to the formation of pomeron loops. 
As the evolution is only leading in colour, such loops cannot be formed 
during the evolution itself. If one e.g.\ studies the collision in a very 
asymmetric frame, where one of the onia carries almost all the available 
energy and the other is almost at rest, 
then the possibility to have multiple collisions is strongly reduced.
Only those pomeron loops are included, which are cut in the specific 
Lorentz frame used for the calculations, which obviously only forms a limited 
set of all possible pomeron loops. It implies that the dipole model is not 
frame independent, and the preferred Lorentz frame is the one where the two 
colliding systems have approximately the same density of dipoles
\footnote{The model is, however, frame independent at the level of one 
pomeron exchange as a consequence of the conformal invariance of the process.}.
This feature clearly limits the rapidity range 
of validity. It is apparent that a frame independent formulation must include 
colour suppressed interactions between the dipoles during the evolution of 
the cascades, but so far it has not been possible to formulate a model, 
which includes saturation effects and is explicitely frame independent.

There is also another related problem with the finite number of colour charges.
The dipole degrees of freedom are 
natural only in the $N_c\rightarrow \infty$ limit. 
Consider for example a system of $r\bar{r}r\bar{r}$ charges.
Using the dipole basis this system the colours can be combined in 
two different ways. 
Obviously, to go beyond large $N_c$ one would need to take into
account quadrupoles, as in the example above, and even
higher multipoles. This makes the colour structure
of the gluon cascade really non-trivial, and one loses 
the picture of a system of dipoles evolving through
dipole splittings in a stochastic process. As the dipole approximation is
so successful in $e^+e^-$-annihilation, it may still be
possible to find a working approximation
using only dipoles. In the example above it may be possible to describe 
the quadrupole field as two dipoles formed by the closest 
charge--anti-charge pairs. 
Such an approximate scheme can be realized by introducing 
a $2\rightarrow 2$ transition vertex in the dipole evolution. 

In this paper we want to discuss ways to improve the dipole description of 
high energy interaction in DIS and hadronic collisions. Some effects of 
energy--momentum conservation were presented in ref.~\cite{Avsar:2005iz}. This constraint 
gives a dynamic cutoff for small dipoles, which strongly reduces some of 
the problems discussed above. The dipole multiplicity is reduced, which 
makes the Monte Carlo simulation much more efficient. The reduced dipole multiplicity
also reduces the effect of saturation, which becomes rather 
small for DIS within the HERA kinematical region. Here we will further extend 
the model presented in \cite{Avsar:2005iz}, including colour suppressed effects related to 
pomeron loops by introducing a $2\rightarrow 2$ transition vertex in the 
dipole evolution. Although not explicitely frame independent, the dependence 
on the Lorentz frame used is here much reduced. 

The coupling of a virtual photon to a $q\bar{q}$ dipole is well
known, but the proton is a much more complicated system.
It was early suggested that semi-hard parton sub-collisions and
minijets are important ingredients in high energy $pp$ collisions,
and responsible for the rising cross section \cite{Cohen-Tannoudji:1982pi,
Mueller:1986ey,Gribov:1984tu,Capella:1992yb}. This picture is 
supported by the successful description of Tevatron data 
\cite{Field:2005qt} using
the P\textsc{ythia} MC, which is based on perturbative parton--parton
collisions \cite{Sjostrand:1987su}. These results encourages us to describe 
high energy $\gamma^*p$ and $pp$
collisions in terms of perturbative dipole--dipole collisions,
and we will in this paper 
also present a simple model for the initial dipole system in a proton. 
The ideas are 
implemented in a computer simulation program, and the results are compared 
with DIS data from HERA and with data from hadron--hadron colliders. 

The paper is organized as follows. In the next section we describe
the dipole picture of high energy collisions in QCD and its relation
to the string picture. In section \ref{sec:jimwlk} we briefly
summarize the alternative approach to high energy QCD, the color glass
condensate, and recent results related to the formation of pomeron
loops in the corresponding evolution equations. In section
\ref{sec:finNcdip} we describe the improvements we have made in the
dipole model as was briefly described in this introduction, followed 
in section \ref{sec:proton} by
a brief discussion of the model of the proton used to obtain
quantitative results. In section \ref{sec:MC} we present some 
details about our Monte Carlo program and how we implement 
the improvements we have made.  Then, in section
\ref{sec:results}, we present the applications of these
improvements and compare our results to experimental data for DIS and $pp$ 
scattering. Finally, in section \ref{sec:conc}, we arrive at our conclusions.

\section{Dipole Picture of QCD}
\label{sec:dipole-intro}

Higher order QCD diagrams are very difficult, and naturally it is not
possible to formulate a quantum mechanical parton cascade as a classical
branching process, including all interference effects. The great success
for parton cascades in $e^+e^-$-annihilation is therefore quite
surprising. DIS is, however, significantly more complicated than
$e^+e^-$-annihilation.

\subsection{Cascades in $e^+e^-$-annihilation}

In $e^+e^-$ the main problem is to calculate the properties of the
final states, while the total cross section is well determined by low
order perturbative QCD calculations.  Although the 1st order matrix
element for gluon emission diverges for soft gluons, the total cross
section is still finite and given by
$\sigma_{tot}\approx\sigma_0\,(1+\alpha_s/\pi)$, where $\sigma_0$ is
the cross section to 0th order in $\alpha_s$. The divergence for soft
gluon emission is compensated by virtual corrections, which can easily
be taken into account by Sudakov form factors. This implies that the
cascade has a probabilistic nature; the emission of one more gluon in
the ordered cascade does not change the total reaction cross section.

The process $e^+e^- \rightarrow q\bar{q}gg$ does factorize in the limit
when one gluon is much softer than the other. In the large $N_c$ limit
also the amplitude for a multi-gluon final state factorizes in the
strongly ordered regime, where one gluon is much softer than the previous
one. $N_c$ is, however, not a big number (even if the non-factorizing
correction terms are of order $1/N_c^2$ and not $1/N_c$). Also the value
of $\alpha_s$ is so large that cascades which are \emph{not} strongly
ordered (and therefore do not factorize), are very important in analyses
of  experimental data.

The result depends quite strongly on the treatment of not well ordered
cascades, where the description depends equally much on physical intuition
as on analytic calculations. A very essential feature is what is called
soft colour interference, which was  mentioned in the introduction. 
A parton with e.g.\ a red colour has 
always a partner carrying anti-red colour charge. The interference between these
two charges implies a suppression for emission of gluons with
wavelengths larger than the separation between the emitters. Thus
the colour charge and its anti-charge partner do not radiate independently,
but must be treated as one unit.

In a fixed Lorentz frame this interference effect can be approximated by
an angular ordering \cite{Mueller:1981ex,Ermolaev:1981cm,Bassetto:1982ma,
Marchesini:1983bm}. This means that emissions from the red and anti-red
partons is restricted to angular cones with opening angles equal to the
angle between the emitters. This angular constraint is not Lorentz
invariant but frame dependent. It also overestimates the emission in some
directions and underestimates it in other, but in such a way that it
averages out to the correct value.

A different approach to this effect is given by the dipole cascade model 
described in \cite{Gustafson:1986db,Gustafson:1988rq}. 
In this model the QCD state is described as a chain of colour dipoles
(with given momentum, energy, and orientation) rather than a chain of
gluons (with momentum, energy, and polarization). This is similar to the
relation between a lattice and its dual lattice. It is interesting to note
that at the end of the cascade this dipole chain gives a smooth transition 
to the string in the Lund fragmentation model \cite{Andersson:1983ia}. In the large $N_c$ limit
the dipoles radiate independently apart from recoils (which may be important
when the emissions are not strongly ordered). The radiation from
the dipole formed by the red and anti-red charges discussed above is
studied in the rest frame of the two partons forming that dipole. Boosting
to e.g.\ the overall rest frame this reproduces the angular ordering, but without
the (unrealistic) sharp angular cutoff. A comparison between the dipole
approximation and the exact 2nd order matrix element is given in
\cite{Andersson:1991he}.

The angular ordering constraint is implemented in the event generators
Herwig \cite{Corcella:2000bw} and Pythia \cite{Sjostrand:2006za}, and
the dipole cascade in the Ariadne event generator
\cite{Lonnblad:1992tz}. They all give very good descriptions of
experimental data from LEP and other $e^+e^-$ colliders, with the
dipole model giving just a slightly better overall $\chi^2$. (The
overall structure of the final state depends strongly on the first
hard emissions, and in all approaches the best agreement is obtained
if these are described by fixed order matrix elements.)

\subsection{Cascades in DIS}

DIS is a more complicated process than $e^+e^-$-annihilation. 
First, there are in DIS two different energy scales, $W^2$
and $Q^2$. Secondly, in DIS both the cross
section and the final state properties are highly nontrivial problems.
Only in the pure DGLAP region, with high $Q^2$ and large $x$, has the
probabilistic description in terms of $k_\perp$-ordered cascades been
really successful, in this case both for cross sections and final state
properties. In the DGLAP region the real emissions are compensated by 
virtual corrections
in a way similar to the time-like cascades in $e^+e^-$-annihilation. Thus the
virtual corrections can also here be treated by Sudakov form factors, and
the cascade contains only the real emissions appearing in the final state.
Thus the DGLAP evolution describes the probability for a given parton
state with a fixed resolution determined by $Q^2$. The total cross section
is determined by the reaction probability between the virtual photon and
the quarks in the cascade. The final state is obtained by adding final
state radiation to the partons in the initial cascade (within angular
ordered regions).

For lower $x$ and $Q^2$ separate approaches have been used to describe the
total cross section and the final state properties. To LL or NLL accuracy
the cross section is determined by the BFKL equation. 
The BFKL evolution can be formulated in different ways. In its
implementation in the Mueller dipole cascade it is not suited to describe
the final state, as the evolution contains a very large number of virtual
dipoles, which do not appear as final state particles. As mentioned,
the BFKL equation
has also the problem that the NLL corrections are so very large.

The presently best description of the final state properties at HERA is
given by the soft radiation model (frequently called the Color Dipole
Model, CDM, and implemented in the Ariadne event generator). In this model
the gluon radiation from the separating colour charges in the kicked out
quark and the proton remnant is described in a way analogous to emission
in $e^+e^-$-annihilation. The CDM model does not predict the cross section, but
only the properties of the final state. It has also the drawback that it
is not solidly founded in perturbative QCD.

The CCFM model \cite{Catani:1990yc,Ciafaloni:1988ur} represents an
interpolation between the DGLAP and BFKL evolutions. In the DGLAP
region the cascade contains, besides the real emissions in the DGLAP
equation, also softer emissions which are treated as final state
radiation in the DGLAP approach. This makes the regions where final
state radiation should be added more complicated, and a description of
final state properties more difficult. The CCFM model is reformulated
and generalized in the Linked Dipole Chain (LDC) model
\cite{Andersson:1995ju}, which is based on a different separation
between initial and final state radiation. Both these models are
implemented in MC event generators,
Cascade\cite{Jung:2000hk,Jung:2001hx} and
LDCMC\cite{Kharraziha:1998dn,Kharraziha:ldcmc} respectively.  The
models have the ambition to describe \emph{both} the cross section and
the final state properties. They both work well with respect to the
cross sections, but none is as successful as the abovementioned CDM
model, when it comes to the properties of the final states.

\subsection{Mueller's Dipole Formulation}
\label{sec:Muellerdip}

The Mueller dipole model \cite{Mueller:1993rr,Mueller:1994jq,Mueller:1994gb}
is formulated in transverse coordinate
space. Such a formulation has the advantage that the transverse
coordinates of the partons are not changed during the evolution. 
This makes it easier to take into account saturation effects
due to multiple scatterings. On the other hand, it is easier to take 
into account energy--momentum conservation in a model formulated
in transverse momentum space. 

In Mueller's model we start with a $q\bar{q}$ pair, heavy enough
for perturbative calculations to be applicable, and calculate
the probability to emit a soft gluon from this pair. Here the 
quark and the antiquark are assumed to follow light-cone trajectories
and the emission of the gluon is calculated in the eikonal 
approximation in which the emitters do not suffer any recoil. Adding the 
contributions to the emission from the quark and the antiquark,
including the interference, 
one obtains the result (for notations, see figure \ref{fig:dipev}) 
\begin{eqnarray}
\frac{d\mathcal{P}}{d\rap}=\frac{\bar{\alpha}}{2\pi}d^2\pmb{z}
\frac{(\pmb{x}-\pmb{y})^2}{(\pmb{x}-\pmb{z})^2 (\pmb{z}-\pmb{y})^2}
\equiv \frac{\bar{\alpha}}{2\pi}d^2\pmb{z} \mathcal{M}(\pmb{x},\pmb{y},\pmb{z}).
\label{eq:dipkernel}
\end{eqnarray}
Here $\pmb{x}$, $\pmb{y}$, and $\pmb{z}$ are two-dimensional vectors
in transverse coordinate space and $\rap$ denotes the rapidity, which 
acts as the time variable in the evolution process.
\FIGURE[t]{
\scalebox{0.7}{\mbox{
\begin{picture}(250,80)(0,5)
\Vertex(10,80){2}
\Vertex(10,0){2}
\Text(5,80)[]{$\pmb{x}$}
\Text(5,0)[]{$\pmb{y}$}
\Line(10,80)(10,0)
\LongArrow(30,40)(60,40)
\Vertex(100,80){2}
\Vertex(100,0){2}
\Vertex(120,50){2}
\Text(95,80)[]{$\pmb{x}$}
\Text(95,0)[]{$\pmb{y}$}
\Text(128,50)[]{$\pmb{z}$}
\Line(100,80)(120,50)
\Line(120,50)(100,0)
\LongArrow(140,40)(170,40)
\Vertex(205,80){2}
\Vertex(225,50){2}
\Vertex(233,30){2}
\Vertex(205,0){2}
\Text(200,80)[]{$\pmb{x}$}
\Text(200,0)[]{$\pmb{y}$}
\Text(233,50)[]{$\pmb{z}$}
\Text(241,30)[]{$\pmb{w}$}
\Line(205,80)(225,50)
\Line(225,50)(233,30)
\Line(233,30)(205,0)
\end{picture}
}}
\caption{\label{fig:dipev} The evolution of the dipole cascade. 
At each step, a dipole can split into two new dipoles 
with decay probability given by \protect\eqref{eq:dipkernel}. }
}
This formula can be interpreted as the emission probability from a dipole
located at $(\pmb{x},\pmb{y})$. In the large $N_c$ limit the gluon 
can be seen as a quark--antiquark pair and the formula above can 
then be interpreted as the decay of the original dipole $(\pmb{x},\pmb{y})$
into two new dipoles, $(\pmb{x},\pmb{z})$ and $(\pmb{z},\pmb{y})$. 
In the same limit further emissions factorize, so that at each step
one has a chain of dipoles where each dipole can decay into 
two new dipoles with the decay probability given by \eqref{eq:dipkernel}.
In this way one obtains a cascade of dipoles which evolve through
dipole splittings, and the number of dipoles grows exponentially with $\rap$. 

We note that the expression above has non-integrable singularities 
at $\pmb{z}=\pmb{x}$ and $\pmb{z}=\pmb{y}$. In numerical calculations 
it is therefore necessary to introduce
a cutoff, $\rho$, such that $(\pmb{x}-\pmb{z})^2,(\pmb{z}-\pmb{y})^2 \geq \rho^2$.
To obtain a meaningful probabilistic interpretation of \eqref{eq:dipkernel} 
(note that $d\mathcal{P}/d\rap$ can become very large for small $\rho$) we also 
need to take into account virtual corrections to the real emissions. 
The effect of these corrections is described by a Sudakov form factor, 
$S=$exp$(-\int d\rap d^2\pmb{z} \cdot d\mathcal{P}/d\rap)$, which should
multiply the splitting probability in \eqref{eq:dipkernel}.

\subsection{Scattering of Dipoles}

In Mueller's model each dipole interacts independently with some target
via two-gluon exchange. In case of onium--onium scattering
each onium evolves into a cascade of dipoles. We let $f_{ij}$ denote the 
imaginary part of the elastic scattering amplitude for a dipole $i$, with 
coordinates $(\pmb{x}_i,\pmb{y}_i)$, in one of the onia and a dipole $j$, 
with coordinates $(\pmb{x}_j,\pmb{y}_j)$, in the other onium. The basic
dipole--dipole scattering amplitude from gluon exchange is given by  
\begin{eqnarray}
f_{ij}=\frac{\alpha_s^2}{8}\biggl[\log\biggl(\frac{(\pmb{x}_i-\pmb{y}_j)^2
(\pmb{y}_i-\pmb{x}_j)^2}{(\pmb{x}_i-\pmb{x}_j)^2(\pmb{y}_i-\pmb{y}_j)^2}\biggr)\biggr]^2.
\label{eq:dipamp}
\end{eqnarray}
In the single pomeron 
approximation the onium--onium amplitude is then
simply given by the sum of the basic dipole--dipole 
amplitudes, $\sum_{ij}f_{ij}$. 

In the dipole model it is also possible to have 
\emph{multiple scatterings}, i.e the simultaneous scatterings
of several dipoles. Assuming that the individual dipole interactions are 
uncorrelated, summing multiple scatterings to all orders exponentiates,
and the total amplitude for a single event is given by 
$1-$exp$(-\sum_{ij}f_{ij})$.
Thus, the expansion of the exponential in a power series corresponds directly
to the contributions from the multiple scattering series, where 
the single pomeron cross section corresponding to the first term in this series
is given by $2\int \langle \sum_{ij}f_{ij}\rangle$.

Consider the scattering of an elementary dipole $(\pmb{x},\pmb{y})$
off some arbitrary target. We denote the scattering matrix 
by $\mathcal{S}(\pmb{x},\pmb{y})$. After one step of evolution in rapidity
the dipole $(\pmb{x},\pmb{y})$ has a chance to split into two 
new dipoles, $(\pmb{x},\pmb{z})$ and $(\pmb{z},\pmb{y})$, through 
the splitting kernel $\mathcal{M}_{\pmb{x}\pmb{y}\pmb{z}}\equiv
\mathcal{M}(\pmb{x},\pmb{y},\pmb{z})$. The evolution of the $S$-matrix
is then given by 
\begin{eqnarray}
\partial_Y\mathcal{S}(\pmb{x},\pmb{y})=\frac{\bar{\alpha}}{2\pi}
\int d^2\pmb{z}\mathcal{M}_{\pmb{x}\pmb{y}\pmb{z}}\{-\mathcal{S}(\pmb{x},\pmb{y})
+\mathcal{S}^{(2)}(\pmb{x},\pmb{z};\pmb{z},\pmb{y})\}.
\end{eqnarray}
The right hand side in this expression simply states that the dipole
can remain as it is, with a reduced probability, $1-\bar{\alpha}/2\pi
\int \mathcal{M}$, or that it can split into two new dipoles, 
$(\pmb{x},\pmb{z})$ and $(\pmb{z},\pmb{y})$, with a probability
density given by \eqref{eq:dipkernel}. If we assume that 
$\mathcal{S}^{(2)}(\pmb{x},\pmb{z};\pmb{z},\pmb{y})=\mathcal{S}(\pmb{x},\pmb{z})
\mathcal{S}(\pmb{z},\pmb{y})$ and rewrite the equation in the 
scattering amplitude $T \equiv 1-\mathcal{S}$, we get
\begin{eqnarray}
\partial_YT(\pmb{x},\pmb{y})=\frac{\bar{\alpha}}{2\pi}
\int d^2\pmb{z}\mathcal{M}_{\pmb{x}\pmb{y}\pmb{z}}\{-T(\pmb{x},\pmb{y})
+ T(\pmb{x},\pmb{z}) + T(\pmb{z},\pmb{y}) - 
T(\pmb{x},\pmb{z})T(\pmb{z},\pmb{y})\}.
\label{eq:theBK}
\end{eqnarray}
This is the so called Balitsky--Kovchegov (BK) equation
\cite{Kovchegov:1999yj,Balitsky:1995ub}.  The assumption that
$\mathcal{S}^{(2)}=\mathcal{S}\mathcal{S}$, corresponds to a mean
field approximation, which can be justified for a large target
nucleus.  As demonstrated in \cite{Mueller:1993rr} the linear part of
\eqref{eq:theBK} reproduces the BFKL equation, while the
inhomogeneous term describes the simultaneous scattering of the two new
dipoles.

The $S$-matrix for a specific scattering event can be 
written as exp$(-\sum_{ij}f_{ij})$. To obtain the physical cross section
one has to perform an average over onium 
configurations, so that $\mathcal{S}=\langle$exp$(-\sum_{ij}f_{ij})\rangle$.
The total cross section is given by $2\int(1-\mathcal{S}(b))$, 
where $b$ denotes the impact parameter, and for 
onium-onium scattering we therefore get
\begin{eqnarray}
\sigma_{tot}=2\int d^2\pmb{b}\langle1-\mathrm{exp}(-\sum_{ij}f_{ij})\rangle.
\label{eq:sigmaonioni}
\end{eqnarray}
For $\gamma^*\gamma^*$ scattering one also needs to convolute 
the averaged amplitude with the virtual photon 
wave functions. The expression in \eqref{eq:sigmaonioni} 
is also what we will use for $\gamma^*p$ 
and $pp$ collisions, where we model the proton as a 
collection of colour dipoles. These points are explained
in greater detail below.

\subsection{Energy--Momentum Conservation}
\label{sec:emcons}

As we saw above, the probability to produce small dipoles diverges
as the size of the dipoles goes to zero. To regulate this divergence
a cutoff, $\rho$, was introduced. Even though this cutoff does not
show up in the cross section (the divergence is canceled 
by virtual corrections, and $\sigma_{\mathrm{tot}}$ approaches a constant 
when $\rho \rightarrow 0$) it must still be kept in a Monte Carlo program. 
A small value of $\rho$, which is needed in order to simulate 
the physics with a good accuracy, will imply that we get very many
small dipoles in the cascade. A small dipole means that 
we have two well localized gluons in the transverse plane, 
and these gluons must then have a correspondingly large 
transverse momentum of the order of the inverse dipole size,
$p_\perp \sim 1/r$. Thus if these small dipoles are 
interpreted as corresponding to real emissions with $p_\perp \sim 1/r$,
then the diverging number of such dipoles would imply the
violation of energy--momentum conservation. 
This suggest that these dipoles should be interpreted as virtual 
fluctuations, which means that the dipole cascade 
will not correspond to the production of exclusive final 
states.

Similarities between Mueller's model and the Linked Dipole Chain (LDC)
model \cite{Andersson:1995ju} were used in ref.~\cite{Avsar:2005iz} to
implement energy conservation in Mueller's model. This removes a
dominant fraction of the virtual emissions. (It does, however, not
remove all virtual emissions. That emissions must satisfy
energy--momentum conservation if they are to be present in real final
states is obviously a necessary condition, but as was discussed in
\cite{Avsar:2005iz}, it is by itself not a sufficient condition.)  The
modified cascade is ordered in both light-cone variables, $p_+$ and
$p_-$, and it was seen that this modification has a rather large
effect on the cascade.  One sees for example that the total number of
dipoles, while still increasing exponentially, is greatly reduced,
which implies that the onset of saturation is delayed. In fact it is
found that in DIS the unitarity effects become quite small within the
HERA energy regime, at least for $Q^2\gtrsim1$~GeV$^2$.  Naturally
saturation is more important for dipole--nucleus or $pp$ scattering.
In particular we will in the following see that saturation effects
have a large influence on $pp$ collisions at the Tevatron.

\section{The JIMWLK Approach}
\label{sec:jimwlk}

\subsection{The Color Glass Condensate}

A different approach to high energy QCD is  
called the Color Glass Condensate (CGC) (for review papers see 
\cite{Iancu:2002xk,Iancu:2003xm,Weigert:2005us}). This is an effective theory
for QCD valid at high gluon densities. Here, the strong gluon fields
present in the high energy particle (which might be a proton, a
large nucleus etc.) emerge due to a classical
random color source, $\rho^a$, and the classical fields 
satisfy the corresponding Yang--Mills 
equations of motion. These random sources are distributed according to a 
weight functional $W[\rho]$. As the particle evolves one proceeds 
by integrating out layers of quantum fields which are added to the 
classical source. This is a renormalization group procedure and 
the weight functional then satisfies a renormalization group equation
which is known as the JIMWLK equation 
\cite{Jalilian-Marian:1997jx,Jalilian-Marian:1997gr,Iancu:2001ad,Weigert:2000gi}. 
The JIMWLK evolution leads
to the saturation of the gluon density\footnote{The growth does not cease completely, 
but it is only logarithmic as opposed to a power-like growth at 
lower energies.} when the field strength 
is of order $1/\alpha_s$. The scale at which the hadron seems to saturate 
is called the saturation momentum, denoted by $Q_s(Y)$. The CGC 
formalism predicts that $Q_s(Y)$ grows exponentially with 
rapidity, defined by $Y=\log(1/x)$ in this case.  

\subsection{The Balitsky-JIMWLK Equations}

When considering a scattering process within the CGC formalism, one
usually thinks of the target as a highly evolved dense particle
which can be described by the weight functional $W[\rho]$, satisfying
the JIMWLK equation. The projectile, on the other hand, is 
usually a simple particle which is not so dense, such as 
an elementary dipole impinging on the target. 
The JIMWLK equation can be written as a Schr\"odinger equation
for the weight functional, $\partial_Y W=H_{JIMWLK}W$, where $H_{JIMWLK}$
denotes the ``JIMWLK Hamiltonian''. Operators corresponding 
to observables are averaged
over with the weight $W[\rho]$, and one may then bring the action of 
$H_{JIMWLK}$ on the operator, instead of on $W[\rho]$ itself. This is
reminiscent of switching from the Schr\"odinger picture (evolution 
of the ``wave function'' $W[\rho]$) to the Heisenberg picture
(evolution of an operator $\mathcal{O}$) in quantum mechanics.   
In particular, if one applies $H_{JIMWLK}$ on $S(\pmb{x},\pmb{y})$, 
the $S$-matrix for the projectile dipole,  an 
infinite hierarchy of equations emerge. This hierarchy of equations is now 
commonly referred to as the Balitsky--JIMWLK (B--JIMWLK) equations, since the 
same set of equations were some years earlier derived by 
Balitsky \cite{Balitsky:1995ub} within a 
different formalism. Taking the large $N_c$ limit\footnote{From now on,
when we talk about the B--JIMWLK equations, we always mean the large 
$N_c$ limit of these equations.} the more complicated 
colour structures disappear, and the equations can be interpreted 
in terms of dipoles evolving according to the discussion 
in section \ref{sec:dipole-intro}. Written in terms of the scattering amplitude
$T(\pmb{x},\pmb{y})=1-S(\pmb{x},\pmb{y})$, the equations 
in this hierarchy can formally be written as
\begin{eqnarray}
\partial_Y \langle T\rangle &=& \mathcal{K}\otimes (
\langle T\rangle - \langle TT\rangle ) \nonumber \\
\partial_Y \langle TT\rangle &=& \mathcal{K}\otimes (
\langle TT\rangle - \langle TTT\rangle )\label{eq:bjimwlk1}\\
\vdots\nonumber
\end{eqnarray}
where $\mathcal{K}$ is the evolution kernel. 
We see here that the equation for $\langle T\rangle$ contains 
a contribution from $\langle TT\rangle$. In turn, the equation for 
$\langle TT\rangle$ contains a term $\langle TTT\rangle$ 
and so on.  

These equations simplify considerably when disregarding target correlations, 
i.e. making a mean field approximation where $\langle TT\rangle = 
\langle T\rangle\langle T\rangle$. As can be seen from \eqref{eq:bjimwlk1}
the hierarchy then boils 
down to a single, closed, nonlinear equation for $\langle T\rangle$, which 
turns out to be none other than the BK equation introduced in section 2. 

\subsection{Inclusion of Pomeron Loops} 

As we saw above the B--JIMWLK hierarchy couples the scattering 
amplitude $\langle T^k\rangle$ to all $\langle T^n\rangle$, with $n \geq k$. There are 
however no contributions from amplitudes $\langle T^n\rangle$ with $n < k$. 
The nonlinear term in the BK 
equation corresponds to pomeron splittings in the projectile;
the projectile dipole splits into two dipoles and each 
of these two couples to the target through a single pomeron giving in total
two pomerons coupling to the target.
However, one can also assume that the target is given the rapidity
increment, and in this case the two pomerons in the target must 
merge into one pomeron, which couples to the single dipole.   
Thus this term also corresponds to the merging 
of two pomerons inside the target (see figure \ref{fig:pomloop}).

\FIGURE[t]{
\scalebox{0.8}{\mbox{
\begin{picture}(400,200)(0,0)
\Photon(260,170)(260,10){3}{8}
\Photon(360,170)(360,10){3}{8}
\Line(250,180)(370,180)
\Line(250,160)(370,160)
\Photon(290,90)(290,10){3}{5}
\Photon(330,90)(330,10){3}{5}
\COval(310,90)(5,55)(0){Black}{Gray}
\COval(310,10)(9,60)(0){Black}{Gray}
\DashLine(250,125)(370,125){3}
\LongArrow(250,40)(250,100)
\rText(375,90)[][l]{$\mathbb{P}$ Merging}
\Photon(10,170)(10,10){3}{8}
\Photon(110,170)(110,10){3}{8}
\Line(0,180)(120,180)
\Line(0,160)(120,160)
\Photon(40,90)(40,10){3}{5}
\Photon(80,90)(80,10){3}{5}
\COval(60,90)(5,55)(0){Black}{Gray}
\COval(60,10)(9,60)(0){Black}{Gray}
\DashLine(0,50)(120,50){3}
\LongArrow(0,140)(0,80)
\rText(125,90)[][l]{$\mathbb{P}$ Splitting}

\end{picture}
}}
\caption
{\label{fig:pomloop} Diagrams for pomeron splittings and mergings. On the left
picture the projectile dipole is evolved, indicated by the down-going 
arrow, and one pomeron is split into two pomerons. The frame 
in which the collision is viewed is indicated by the horizontal 
dashed line. On the picture to the right, the target, the gray blob at
the bottom, is evolved and two pomerons merge into one pomeron which 
couple to the projectile.}
}

We therefore see that the B--JIMWLK equations describe either
pomeron mergings, when the target is evolved, or
pomeron splittings, in case the projectile is 
evolved, but not both. Thus, even though 
the CGC formalism correctly describes saturation 
effects, it nevertheless misses some essential physics
as it cannot account for pomeron splittings. What is
actually absent is gluon number fluctuations. Indeed, 
in the CGC approach the small-$x$ gluons are radiated
from the classical colour source $\rho$, but are 
themselves not allowed to split. They rather
get absorbed into $W[\rho]$, and act as sources
for gluons with even smaller $x$, as the evolution proceeds.
The effects of fluctuations were demonstrated 
in numerical studies by Salam \cite{Mueller:1996te}, and it is known
that they are important for the approach
towards the unitarity limit \cite{Iancu:2004es,Iancu:2003zr}. We note 
that these fluctuations are correctly taken into
account in the dipole model, and in a Monte Carlo
program based on it, as demonstrated by Salam. 

Ever since it was realized that the B--JIMWLK equations
are not complete, there has been a lot of effort to construct a
model which contains both pomeron mergings and splittings, and,  
through iterations, pomeron loops. This has been formulated in 
the large $N_c$ limit \cite{Iancu:2004iy,Mueller:2005ut,Levin:2005au} where the dipole
model has been used to add pomeron splittings to the B--JIMWLK equations
in the dilute region. The extension to the dense region 
is then obtained by simply adding the remaining 
terms arising from the large $N_c$ version of the B--JIMWLK
hierarchy. The main principle is that the two kinds of 
pomeron interactions (splittings and mergings) are important
in different, well separated, kinematical regions. The equations
obtained in this way give the correct expressions 
in the two limits (dense and dilute
systems), but it is not very clear how well they work in an 
intermediate region. The new equation for $\langle TT\rangle$  receives 
a contribution also from  $\langle T\rangle$ and it can be written as 
\begin{eqnarray}
\partial_Y \langle TT\rangle = \mathcal{K}\otimes (
\langle TT\rangle - \langle TTT\rangle ) + \mathscr{F}\cdot
\langle T\rangle,
\label{eq:pomloopev}
\end{eqnarray}
where $\mathscr{F}$ is a quite complicated expression
describing the fluctuations in the target (or saturation
effects in the projectile). 

\section{Finite $N_c$ Effects in Dipole Evolution}
\label{sec:finNcdip}

In this section we want to discuss what improvements can be made
in order to obtain a more complete picture of high energy
evolution using the dipole degrees of freedom. In a formalism where both the
projectile and the target are considered within the dipole 
picture, the missing piece is saturation 
effects rather than fluctuations, which are fully accounted for
in the dipole model. 

In many approaches the dipoles in a cascade are treated as independent 
and without a specified direction.
An important feature in our formalism is that our cascade consists of a
\emph{chain} of dipoles which are all connected to each other through
the gluons. This chain has also a 
specified \emph{direction}, with each dipole oriented from colour charge 
to anti-charge. Such a chain can only end in a quark or an antiquark. In this 
picture one therefore cannot
simply take two arbitrary dipoles and merge them into one dipole, leaving 
loose ends behind. It is also necessary to specify how these ends
afterwards get reconnected to other dipoles in the systems. We therefore
begin this section by a discussion of colour structures, and a motivation
why it is important, before
engaging into the problems due to the finite number of colours.
We end the section with a comparison between our formalism and other
approaches to include colour suppressed effects in the dipole cascade
formalism.

\subsection{Colour Structures}
\label{sec:colstruct1}

In onium--onium scattering it is assumed that the probability for a
dipole--dipole sub-collision is independent of the remaining dipoles in
the cascades.  The exchange of a gluon implies that the intermediate
state corresponds to a recoupling of the colour flow, as is shown in
fig.~\ref{fig:dipint}.  This interaction actually corresponds to the
coherent sum of four different Feynman diagrams, illustrated in
fig.~\ref{fig:colexch}.  Note in particular that in the dipole
formalism a dipole is a colour singlet, i.e. a coherent sum of
$r\bar{r}$, $b\bar{b}$, and $g\bar{g}$. Therefore the diagrams in
figs.~\ref{fig:colexch}c and \ref{fig:colexch}d have the same weight
as those in figs.~\ref{fig:colexch}a and \ref{fig:colexch}b. Summing
and averaging over colours they are all of order $\alpha_s$, and thus
formally colour suppressed compared to the dipole splitting vertex in
\eqref{eq:dipkernel}, which is proportional to $\bar{\alpha}=N_c\,
\alpha_s / \pi$.

\FIGURE[t]{
\scalebox{1.0}{\mbox{
\begin{picture}(240,275)(0,5)
\Oval(10,215)(20,10)(0)
\Vertex(10,223){2}
\Text(0,223)[r]{$q$}
\Vertex(10,207){2}
\Text(0,207)[r]{$\bar{q}$}
\LongArrow(30,215)(60,215)
\Vertex(80,275){2}
\Text(75,275)[r]{$q$}
\Vertex(87,250){2}
\Vertex(92,225){2}
\Text(87,225)[r]{$x_i$}
\Vertex(92,200){2}
\Text(87,200)[r]{$y_i$}
\Vertex(87,175){2}
\Vertex(80,150){2}
\Text(75,150)[r]{$\bar{q}$}
\ArrowLine(80,275)(87,251)
\ArrowLine(87,249)(92,226)
\ArrowLine(92,224)(92,201)
\ArrowLine(92,199)(87,176)
\ArrowLine(87,174)(80,151)
\Oval(220,215)(20,10)(0)
\Text(237,223)[r]{$\bar{q}$}
\Vertex(220,207){2}
\Vertex(220,223){2}
\Text(237,207)[r]{$q$}
\LongArrow(200,215)(170,215)
\Vertex(150,275){2}
\Text(155,275)[l]{$\bar{q}$}
\Vertex(143,245){2}
\Text(148,245)[l]{$y_j$}
\Vertex(137,200){2}
\Text(142,200)[l]{$x_j$}
\Vertex(150,155){2}
\Text(155,155)[l]{$q$}
\ArrowLine(143,246)(150,275)
\ArrowLine(137,201)(143,244)
\ArrowLine(150,156)(137,199)
\LongArrow(114,155)(114,125)
\Vertex(80,125){2}
\Text(75,125)[r]{$q$}
\Vertex(87,100){2}
\Vertex(92,75){2}
\Text(87,75)[r]{$x_i$}
\Vertex(92,50){2}
\Text(87,50)[r]{$y_i$}
\Vertex(87,25){2}
\Vertex(80,0){2}
\Text(75,0)[r]{$\bar{q}$}
\ArrowLine(80,125)(87,101)
\ArrowLine(87,99)(92,76)
\ArrowLine(92,76)(143,96)
\ArrowLine(143,96)(150,125)
\Vertex(150,125){2}
\Text(155,125)[l]{$\bar{q}$}
\Vertex(143,95){2}
\Text(148,95)[l]{$y_j$}
\Vertex(137,50){2}
\Text(142,50)[l]{$x_j$}
\Vertex(150,5){2}
\Text(155,5)[l]{$q$}
\ArrowLine(92,49)(87,26)
\ArrowLine(87,24)(80,1)
\ArrowLine(150,6)(137,49)
\ArrowLine(137,49)(92,49)
\end{picture}
}}
\caption{\label{fig:dipint} Symbolic picture showing the interaction
of two onia via a single sub-collision. The interaction between the 
dipoles $(\pmb{x}_i,\pmb{y}_i)$ and $(\pmb{x}_j, \pmb{y}_j)$ leads 
to a recoupling of the colour flow, with strength $f_{ij}$ given 
by \protect\eqref{eq:dipamp}.}
}

\FIGURE[t]{
\scalebox{1.0}{\mbox{
\begin{picture}(370,165)(0,5)
\Vertex(10,160){2}
\Vertex(10,120){2}
\Vertex(60,160){2}
\Vertex(60,120){2}
\ArrowLine(10,160)(10,120)
\ArrowLine(60,120)(60,160)
\Photon(20,155)(50,125){2}{4}
\Text(38,150)[]{$r\bar{b}$}
\LongArrow(28,137)(38,126)
\LongArrow(35,110)(35,80)
\Vertex(10,60){2}
\Vertex(10,20){2}
\Vertex(60,60){2}
\Vertex(60,20){2}
\ArrowLine(10,60)(60,60)
\ArrowLine(60,20)(10,20)
\Text(35,0)[]{$\pmb{a})$}
\Text(0,160)[]{$r$}
\Text(0,120)[]{$\bar{r}$}
\Text(70,160)[]{$\bar{b}$}
\Text(70,120)[]{$b$}
\Text(0,60)[]{$b$}
\Text(0,20)[]{$\bar{r}$}
\Text(70,60)[]{$\bar{b}$}
\Text(70,20)[]{$r$}
\Vertex(110,160){2}
\Vertex(110,120){2}
\Vertex(160,160){2}
\Vertex(160,120){2}
\ArrowLine(110,160)(110,120)
\ArrowLine(160,120)(160,160)
\Photon(120,125)(150,155){2}{4}
\Text(135,150)[]{$b\bar{r}$}
\LongArrow(133,130)(143,138)
\LongArrow(135,110)(135,80)
\Vertex(110,60){2}
\Vertex(110,20){2}
\Vertex(160,60){2}
\Vertex(160,20){2}
\ArrowLine(110,60)(160,60)
\ArrowLine(160,20)(110,20)
\Text(135,0)[]{$\pmb{b})$}
\Text(100,160)[]{$r$}
\Text(100,120)[]{$\bar{r}$}
\Text(170,160)[]{$\bar{b}$}
\Text(170,120)[]{$b$}
\Text(100,60)[]{$r$}
\Text(100,20)[]{$\bar{b}$}
\Text(170,60)[]{$\bar{r}$}
\Text(170,20)[]{$b$}
\Vertex(210,160){2}
\Vertex(210,120){2}
\Vertex(260,160){2}
\Vertex(260,120){2}
\ArrowLine(210,160)(210,120)
\ArrowLine(260,120)(260,160)
\Photon(220,160)(250,160){2}{4}
\Text(235,150)[]{$r\bar{b}$}
\LongArrow(231,165)(242,165)
\LongArrow(235,110)(235,80)
\Vertex(210,60){2}
\Vertex(210,20){2}
\Vertex(260,60){2}
\Vertex(260,20){2}
\ArrowLine(210,60)(260,60)
\ArrowLine(260,20)(210,20)
\Text(235,0)[]{$\pmb{c})$}
\Text(200,160)[]{$r$}
\Text(200,120)[]{$\bar{r}$}
\Text(270,160)[]{$\bar{r}$}
\Text(270,120)[]{$r$}
\Text(200,60)[]{$b$}
\Text(200,20)[]{$\bar{r}$}
\Text(270,60)[]{$\bar{b}$}
\Text(270,20)[]{$r$}
\Vertex(310,160){2}
\Vertex(310,120){2}
\Vertex(360,160){2}
\Vertex(360,120){2}
\ArrowLine(310,160)(310,120)
\ArrowLine(360,120)(360,160)
\Photon(320,120)(350,120){2}{4}
\Text(335,130)[]{$b\bar{r}$}
\LongArrow(331,115)(342,115)
\LongArrow(335,110)(335,80)
\Vertex(310,60){2}
\Vertex(310,20){2}
\Vertex(360,60){2}
\Vertex(360,20){2}
\ArrowLine(310,60)(360,60)
\ArrowLine(360,20)(310,20)
\Text(335,0)[]{$\pmb{d})$}
\Text(300,160)[]{$r$}
\Text(300,120)[]{$\bar{r}$}
\Text(370,160)[]{$\bar{r}$}
\Text(370,120)[]{$r$}
\Text(300,60)[]{$r$}
\Text(300,20)[]{$\bar{b}$}
\Text(370,60)[]{$\bar{r}$}
\Text(370,20)[]{$b$}
\end{picture}
}}
\caption{\label{fig:colexch} Diagrams for dipole--dipole scattering. Each interaction
implies a recoupling of the colour flow and the square of the sum of the 
four diagrams give $f_{ij}$ in \eqref{eq:dipamp}.} 
}

We see that the result of the interactions in figs.~\ref{fig:dipint} and
\ref{fig:colexch} is two new, directed and uniquely specified,
dipole chains. The colour end from one initial dipole chain is
connected to the anti-colour end from the other initial chain. 
There are actually two good reasons to keep track of the dipole orientations.
First we note that given the orientation of the colliding 
dipoles the final dipoles are uniquely determined. There is only one 
possible way to connect the four involved gluons and only one possible
orientation for the new dipoles. Thus keeping track of the orientation 
actually simplifies the formalism, as knowing which
end of the dipole is the colour and which is the anti-colour reduces the 
number of contributing Feynman diagrams. Secondly we have the ambition 
to include  analyses of exclusive final states in future work,
and it is clearly necessary to keep track of the orientation 
of the dipole chains when we want to add final state radiation and 
hadronization.

Multiple
dipole--dipole sub-collisions give more complicated final states, as
illustrated in figs.~\ref{fig:multchain}a and \ref{fig:multchain}b.
When dipole 1 scatters against dipole 3 and dipole 2 against dipole 4,
as shown in fig.~\ref{fig:multchain}a, the result includes an isolated
dipole loop in the center. If instead dipole 1 scatters against
dipole 4 and dipole 2 against dipole 3, as in
fig.~\ref{fig:multchain}b, the result is two dipole chains, each
connecting the two ends from one of the initial incoming chains.
The lower figures give schematic pictures of the resulting 
dipole chains. Here the projectile and target remnants move to
the right and left respectively. The dipole chains are stretched
between these remnants and the gluons which have participated in the hard
sub-collisions.

\FIGURE[t]{
  (a)\epsfig{file=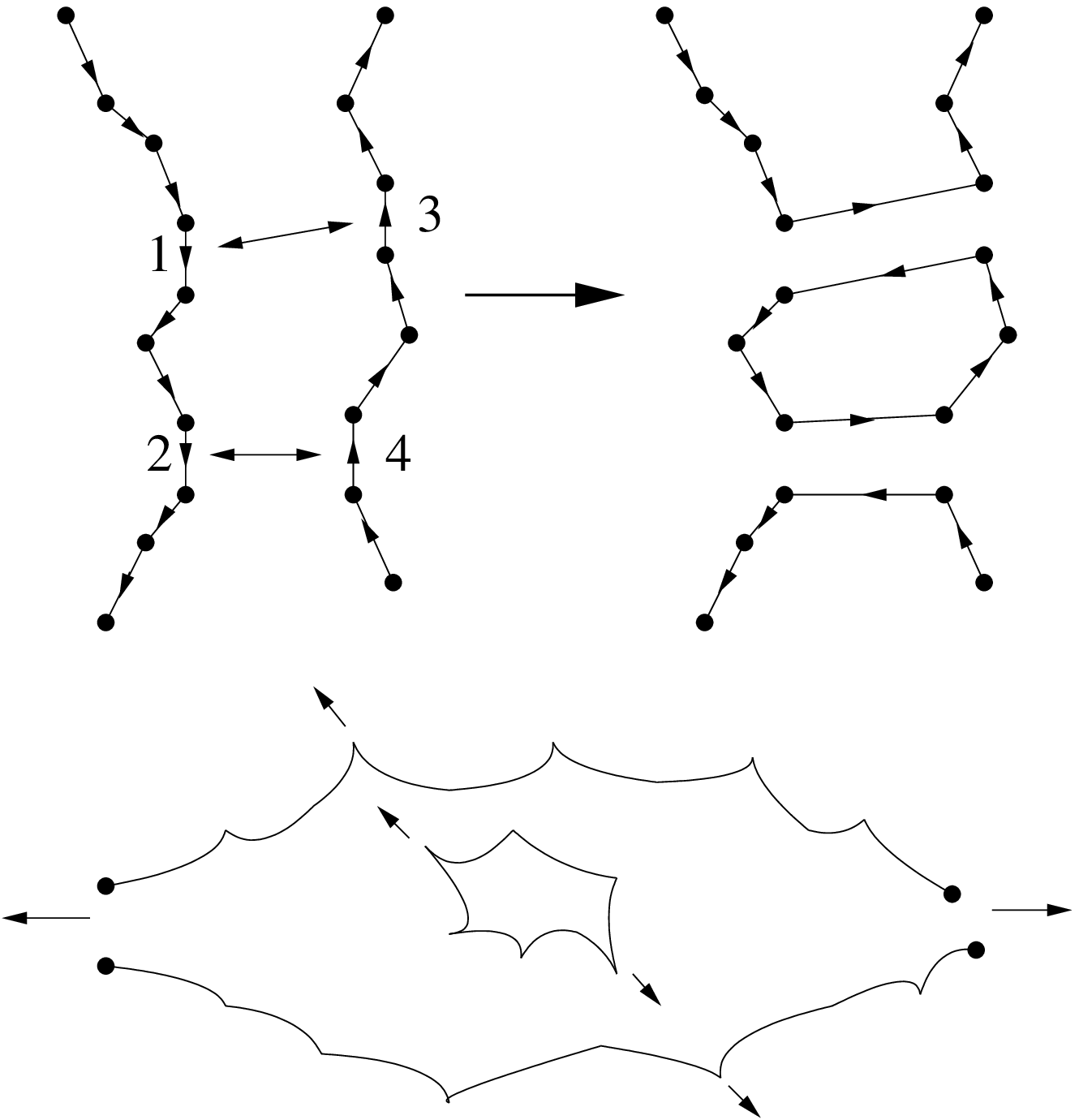,width=6cm}\hfill
  (b)\epsfig{file=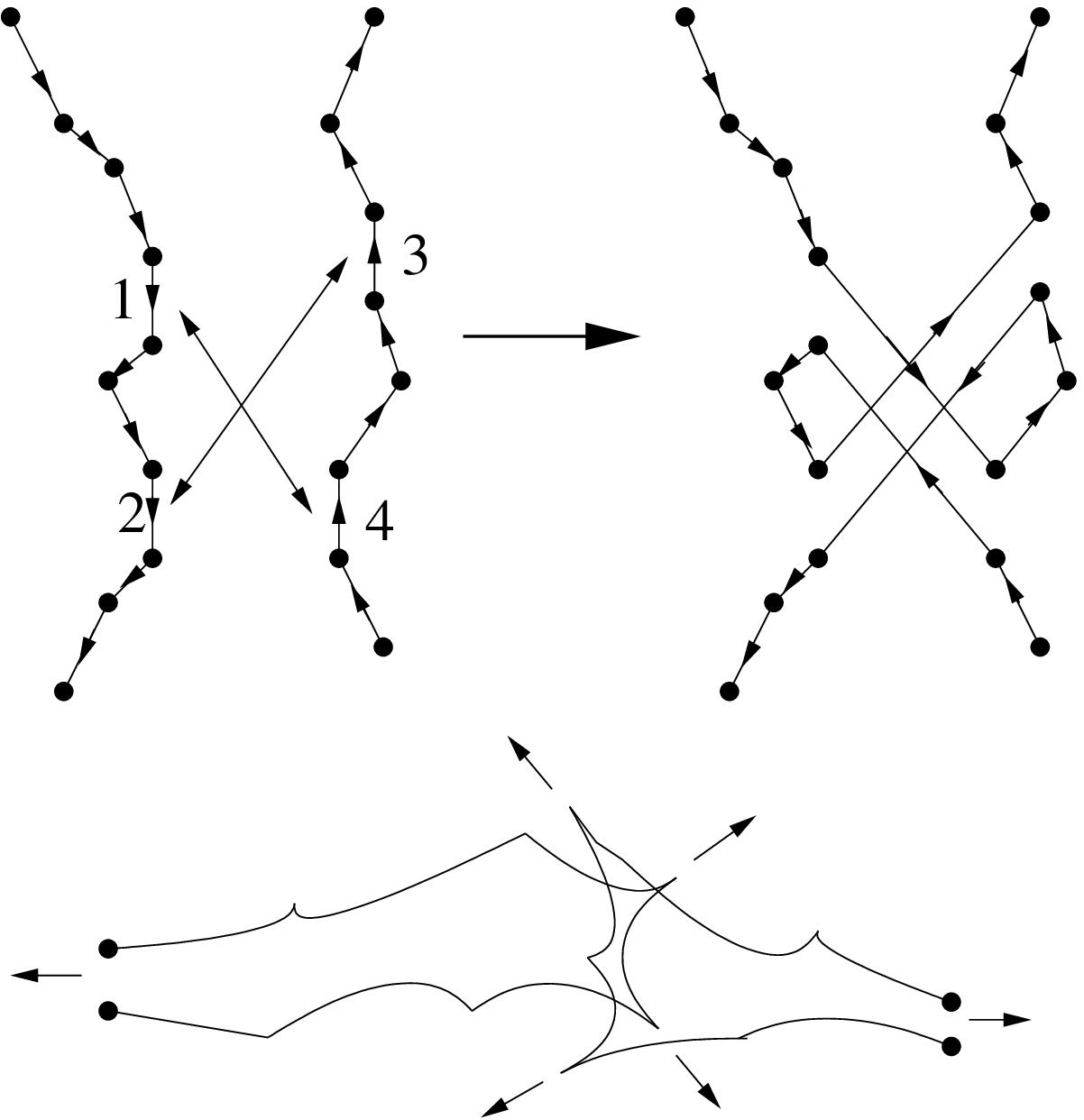,width=6cm}

  \caption{\label{fig:multchain} The colour structure arising from two 
    sub-collisions between the right- and left-moving onia. (a) The
    final configuration involves an isolated dipole loop together with
    two chains, each connecting the initial quark (antiquark) in the
    right moving onia with the initial antiquark (quark) in the left
    moving onia.  (b) The final configuration obtained when dipole 1 collides
    with 4 and 2 with 3. The result is two ``entangled'' chains.
    The lower part of the pictures give a schematic view of the resulting 
    dipole chains, with the projectile and target remnants moving to
    the right and left respectively. Gluons participating in the hard
    sub-collisions are also indicated by arrows.}  }

\FIGURE[t]{
\scalebox{0.8}{\mbox{
\begin{picture}(400,180)(0,5)
\Photon(260,170)(260,10){3}{5}
\Photon(390,170)(390,10){3}{5}
\Line(250,180)(400,180)
\Line(250,160)(400,160)
\Line(250,0)(400,0)
\Line(250,20)(400,20)
\SetColor{Red}
\Photon(260,140)(390,140){3}{5}
\Photon(260,120)(390,120){3}{5}
\SetColor{Black}
\Photon(303,118)(303,58){3}{3}
\Photon(347,122)(347,62){3}{3}
\SetColor{Blue}
\Photon(262,100)(392,100){3}{5}
\Photon(258,80)(388,80){3}{5}
\SetColor{Magenta}
\Photon(305,110)(345,110){3}{3}
\Photon(305,90)(345,90){3}{3}
\Photon(305,70)(345,70){3}{3}
\SetColor{Red}
\Photon(260,40)(390,40){3}{5}
\Photon(260,60)(390,60){3}{5}
\SetColor{Red}
\CArc(50,170)(10,90,270)
\Line(50,180)(150,180)
\Line(50,160)(80,160)
\SetColor{Blue}
\Line(90,160)(150,160)
\Line(90,160)(90,145)
\Line(90,145)(100,145)
\Line(100,140)(90,140)
\Line(90,140)(90,130)
\Line(90,130)(150,130)
\SetColor{Magenta}
\Line(90,120)(120,120)
\Line(120,120)(120,110)
\Line(120,110)(114,110)
\Line(114,105)(120,105)
\Line(120,105)(120,95)
\Line(120,95)(114,95)
\Line(114,90)(120,90)
\Line(120,90)(120,80)
\Line(114,80)(120,80)
\Line(114,75)(120,75)
\Line(120,75)(120,65)
\Line(114,65)(120,65)
\Line(114,60)(120,60)
\Line(120,60)(120,50)
\Line(120,50)(90,50)
\SetColor{Magenta}
\Line(90,110)(90,120)
\Line(90,110)(96,110)
\Line(96,105)(90,105)
\Line(90,105)(90,95)
\Line(90,95)(96,95)
\Line(96,90)(90,90)
\Line(90,90)(90,80)
\Line(96,80)(90,80)
\Line(96,75)(90,75)
\Line(90,75)(90,65)
\Line(96,65)(90,65)
\Line(96,60)(90,60)
\Line(90,60)(90,50)
\SetColor{Blue}
\Line(130,120)(130,105)
\Line(130,120)(150,120)
\Line(130,105)(135,105)
\Line(130,100)(135,100)
\Line(130,100)(130,85)
\Line(130,85)(135,85)
\Line(130,80)(135,80)
\Line(130,80)(130,70)
\Line(130,70)(135,70)
\Line(130,65)(135,65)
\Line(130,65)(130,50)
\Line(130,50)(150,50)
\SetColor{Blue}
\Line(150,40)(90,40)
\Line(90,40)(90,35)
\Line(90,35)(100,35)
\Line(90,30)(100,30)
\Line(90,30)(90,20)
\Line(90,20)(150,20)
\SetColor{Red}
\Line(50,0)(150,0)
\CArc(50,10)(10,90,270)
\Line(50,20)(80,20)
\Line(80,160)(80,150)
\Line(80,150)(70,150)
\Line(70,145)(80,145)
\Line(80,145)(80,135)
\Line(80,135)(70,135)
\Line(80,130)(70,130)
\Line(80,130)(80,110)
\Line(80,110)(80,95)
\Line(80,95)(70,95)
\Line(80,90)(70,90)
\Line(80,90)(80,70)
\Line(80,70)(70,70)
\Line(80,65)(70,65)
\Line(80,65)(80,40)
\Line(80,40)(70,40)
\Line(80,35)(70,35)
\Line(80,35)(80,20)
\SetColor{Black}
\Photon(0,170)(40,170){2}{4}
\Photon(0,10)(40,10){2}{4}
\Photon(85,160)(85,20){3}{6}
\Photon(85,125)(150,125){3}{3}
\Photon(85,45)(150,45){3}{3}
\Photon(125,124)(125,44){3}{3}
\end{picture}
}}
\caption
{\label{fig:diptopom} A high energy collision showing the partonic
  sub-collisions inside the resolved photons. This figure shows the
  case of two sub-collisions and in the left figure there is a loop of
  dipoles at the center. To the right is the corresponding elastic
  diagram which shows the exchange of two pomerons.}  }

\subsection{Effects of Finite $N_c$}

\FIGURE[t]{
  \epsfig{file=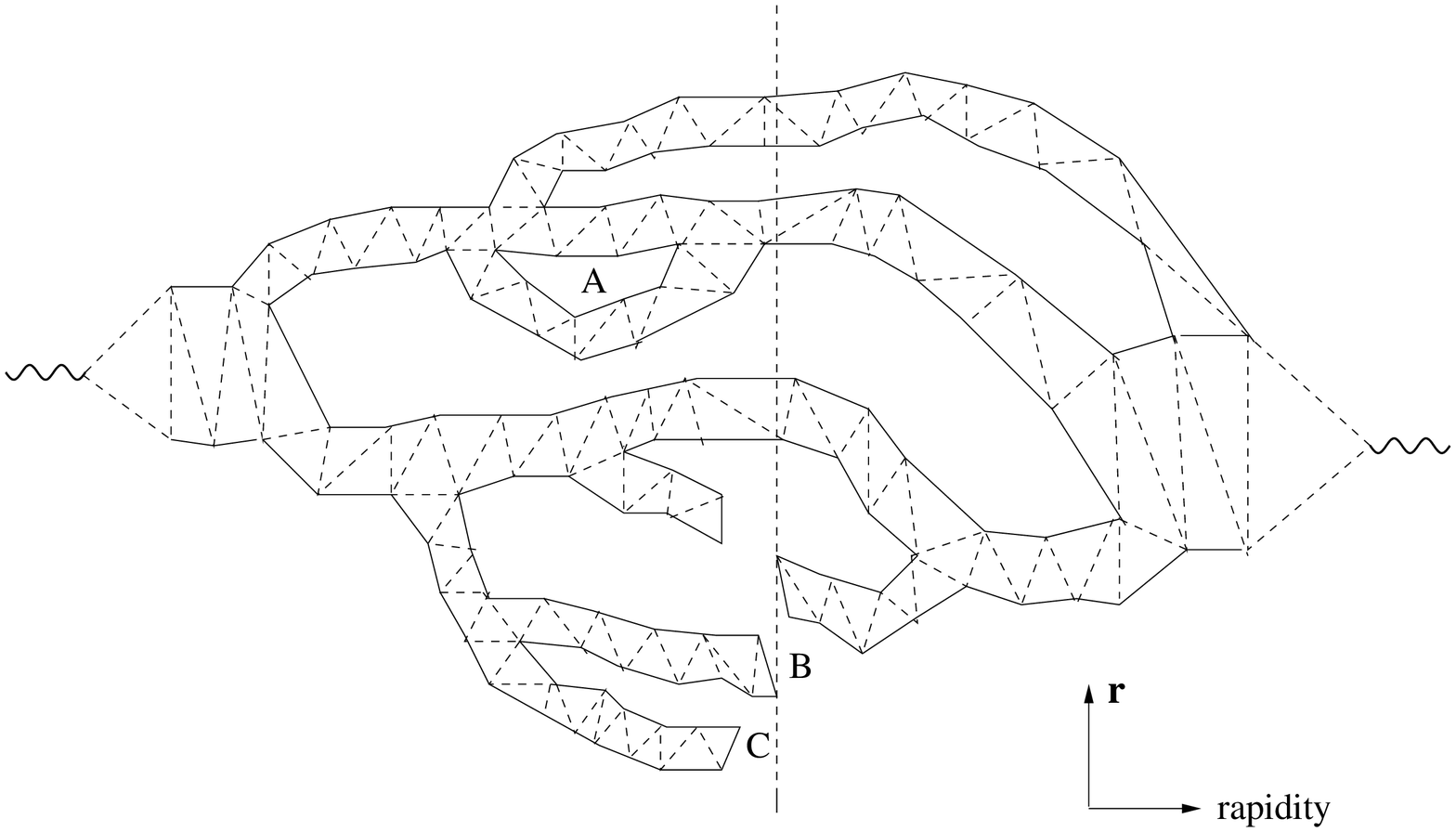,width=10cm}
  \caption{\label{fig:multcoll} Collision of two dipole cascades 
    in $\pmb{r}$-rapidity space. The dashed vertical line symbolizes
    the Lorentz frame in which the collision is viewed. The
    dipole splitting vertex can result in the formation of different dipole
    branches, and loops are formed
    due to multiple sub-collisions. The loop denoted by $A$ can be
    formed via a dipole swing, which is further illustrated in fig
    \ref{fig:dipswing}.  } }

As discussed in the introduction there are two different effects related 
to the finite number of colours.
The first problem is due to the fact that the amplitude for a dipole--dipole 
collision is proportional to $\alpha_s$, and therefore formally colour 
suppressed compared to the dipole splitting process proportional
to $\bar{\alpha}=N_c \alpha_s / \pi$. 
Thus, while one takes into account colour-suppressed effects and saturation
due to multiple dipole--dipole sub-collisions, the evolution itself does not 
contain such effects. The multiple collision events can lead 
to the formation of colour loops, as illustrated in 
fig.~\ref{fig:multchain}a, or to pomeron loops in the elastic amplitude
as shown in fig.~\ref{fig:diptopom}. 
Fig.~\ref{fig:multcoll} shows an example with a more complicated event, 
where three dipole--dipole sub-collisions result in the formation of two loops. 
There is also one loop 
formed totally within one of the cascades, indicated by the letter $A$.
Such a loop cannot be formed within Mueller's initial formalism,
in which only dipole splitting is included within the cascade.
It could, however, have been included if the reaction had been studied in a 
different Lorentz frame. We see that in order to achieve a boost invariant
formalism we must allow dipoles to combine in the cascade.
 We note that 
the new terms that were included in the B--JIMWLK equations, 
discussed above, are also formally colour suppressed and
are essential in order to obtain a frame independent formalism.

We should however point out that there is also another frame 
dependent effect, which is more kinematic in origin. For evolution 
with a finite cutoff, $\rho \neq 0$, frame independence is only 
approximative, even in the one pomeron approximation (only one branch
coupling to the target). In our case 
we have a dynamical cutoff, $\rho(\Delta y)$ (see section \ref{sec:MC}),
and in our scheme every new branch takes away energy. This means that in 
a cascade with many branches the energy in each individual branch 
is reduced. We note that a branch can only be realized if it interacts 
with the target and branches which do not interact have to be regarded 
as virtual (such examples are shown in fig.~\ref{fig:multcoll} 
where the branches marked B and C do not couple to the target). 
These branches should consequently be removed from the cascade 
and in the corresponding final state they should be replaced by their 
earlier ancestors. However, 
as our constraint from energy-momentum conservation also includes the 
fractions needed to evolve the non-interacting branches the effect is 
somewhat overestimated. Therefore we do not expect to find complete 
frame independence, but we will see in the following that the 
frame dependence is indeed very small.         
 

The second problem is due to the possibility that two dipoles can have 
the same colour. The two charges and their corresponding anti-charges then 
form a colour quadrupole, which cannot be described as two independent dipoles.

\subsubsection{Gluon Exchange}

The two dipole sub-collisions in fig.~\ref{fig:multchain}a, which both are 
due to single gluon exchange, lead to a recoupling of the dipole chains and
a closed dipole
loop. Visualized in a different Lorentz frame this process
must be interpreted as the result of gluon exchange between two dipoles
in the cascade. It then corresponds to the replacement of two 
dipoles with two new dipoles within the evolution of the cascade. 
This process has been called a 
``dipole swing'' and is illustrated in fig.~\ref{fig:swing}. 
As it represents the dipole--dipole scattering cross section in 
\eqref{eq:dipamp} it ought to be proportional to $\alpha_s^2/8$, and therefore
effectively suppressed.
We ought to
point out that fig.~\ref{fig:swing} is only a schematic picture showing how 
the dipoles are connected to each other, and does not represent the transverse
size of the dipoles. In fact, the swing process is 
more likely to replace two dipoles with two smaller dipoles, as discussed below.

\FIGURE[t]{
\scalebox{1.0}{\mbox{
\begin{picture}(250,200)(0,5)
\Oval(10,110)(20,10)(0)
\Vertex(10,118){2}
\Text(0,118)[r]{$q$}
\Vertex(10,102){2}
\Text(0,102)[r]{$\bar{q}$}
\LongArrow(30,110)(60,110)
\Vertex(80,200){2}
\Text(75,200)[r]{$q$}
\Vertex(84,180){2}
\Vertex(88,160){2}
\Text(93,160)[l]{$x_1$}
\Vertex(92,140){2}
\Text(97,140)[l]{$y_1$}
\Vertex(94,120){2}
\Vertex(94,100){2}
\Vertex(92,80){2}
\Text(97,80)[l]{$x_2$}
\Vertex(88,60){2}
\Text(93,60)[l]{$y_2$}
\Vertex(84,40){2}
\Vertex(80,20){2}
\Text(75,20)[r]{$\bar{q}$}
\ArrowLine(80,200)(84,181)
\ArrowLine(84,179)(88,161)
\ArrowLine(88,159)(92,141)
\ArrowLine(92,139)(94,121)
\ArrowLine(94,119)(94,101)
\ArrowLine(94,99)(92,81)
\ArrowLine(92,79)(88,61)
\ArrowLine(88,59)(84,41)
\ArrowLine(84,39)(80,21)
\LongArrow(120,110)(150,110)
\Vertex(200,200){2}
\Text(195,200)[r]{$q$}
\Vertex(204,180){2}
\Vertex(208,160){2}
\Text(213,160)[l]{$x_1$}
\Vertex(212,140){2}
\Text(217,140)[l]{$y_1$}
\Vertex(214,120){2}
\Vertex(214,100){2}
\Vertex(212,80){2}
\Text(217,80)[l]{$x_2$}
\Vertex(208,60){2}
\Text(213,60)[l]{$y_2$}
\Vertex(204,40){2}
\Vertex(200,20){2}
\Text(195,20)[r]{$\bar{q}$}
\ArrowLine(200,199)(204,181)
\ArrowLine(204,179)(208,161)
\Curve{(185,110)(186,115)(187,120)(190,130)(195,140)(201,150)(208,159)}
\Curve{(185,110)(186,105)(187,100)(190,90)(195,80)(201,70)(208,61)}
\LongArrow(185,111)(185,109)
\ArrowLine(212,139)(214,121)
\ArrowLine(214,119)(214,101)
\ArrowLine(214,99)(212,81)
\Curve{(200,110)(201,106)(202,100)(205,90)(212,81)}
\Curve{(200,110)(201,114)(202,120)(205,130)(212,139)}
\LongArrow(200,109)(200,111)
\ArrowLine(208,59)(204,41)
\ArrowLine(204,39)(200,21)
\end{picture}
}}
\caption{\label{fig:swing} Schematic picture of a dipole swing. 
  The two dipoles $(\pmb{x}_1,\pmb{y}_1)$ and
  $(\pmb{x}_2,\pmb{y}_2)$ are transformed into two new dipoles
  $(\pmb{x}_1,\pmb{y}_2)$ and $(\pmb{x}_2,\pmb{y}_1)$ after a
  recoupling of the colour flow. The initial chain of dipoles is
  replaced by a new chain stretching between the original $q\bar{q}$
  pair and a loop of dipoles.}  }

Including the dipole swing it is in fact possible to generate any kind of colour
loop. Thus all loops formed when the expanding ``tentacles'' in 
fig.~\ref{fig:multcoll}
join can be generated by the original dipole splitting process together 
with the dipole swing. This is illustrated in fig.~\ref{fig:dipswing},
which shows how a dipole splitting process in the evolution towards the left
can also be visualized as a pomeron fusion process generated by the dipole
swing, when the process is evolved in the opposite direction.

\FIGURE[t]{
  \epsfig{file=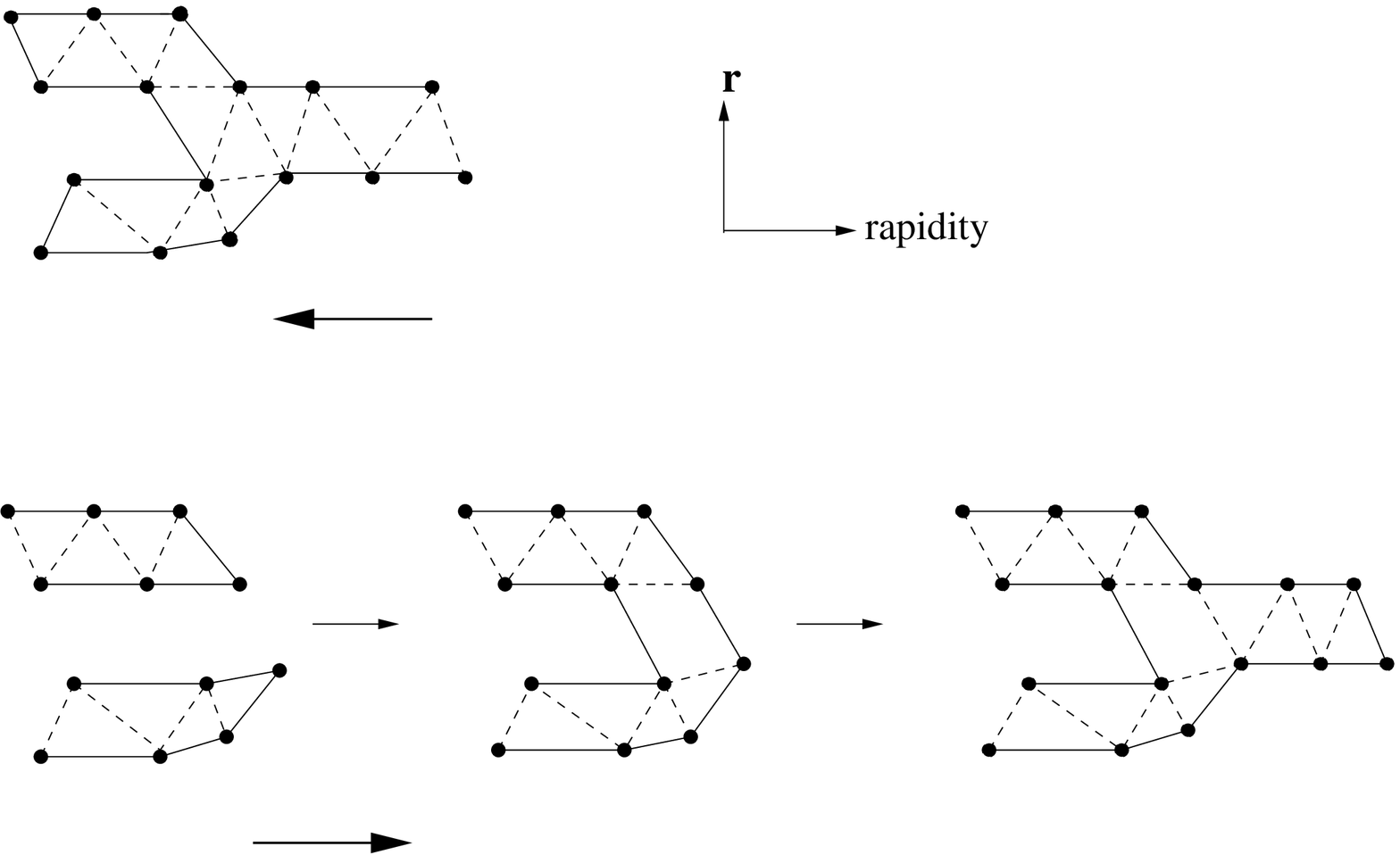,width=10cm}
  \caption{\label{fig:dipswing} Evolution of dipoles in $\pmb{r}$-rapidity space. 
    Going to the left we have evolution through dipole splittings and
    one dipole chain splits into two, corresponding to a
    $1\longrightarrow 2$ pomeron splitting.  If instead we evolve to
    the right then the two chains can be combined by a swing and only
    one of the chains continue evolving, corresponding to a
    $2\longrightarrow 1$ pomeron merging.}  }

\subsubsection{Colour Multipoles}

As mentioned above a second effect of finite $N_c$ is the formation
of colour quadrupoles and higher multipoles. In fact, it can be seen 
in the complete version of the B--JIMWLK equations that 
more complicated colour structures appear at each step of the 
evolution. Obviously these complicated colour structures 
imply that one loses the simple picture of a system
of dipoles, which evolve through simple splittings.
Nevertheless, in view of the success of the time-like dipole cascades in
$e^+e^-$-annihilation it may be possible to find a working approximation
within the dipole framework also in this case. 
We may then try to approximate a quadrupole as two dipoles where those
formed by the closest colour--anti-colour pairs should
dominate. This means that we allow for 
a colour recoupling, in which two dipoles with
coordinates $(\pmb{x}_1,\pmb{y}_1)$ and $(\pmb{x}_2,\pmb{y}_2)$, can be
transformed into two new dipoles with coordinates
$(\pmb{x}_1,\pmb{y}_2)$ and $(\pmb{x}_2,\pmb{y}_1)$.

We note that the result of this process also corresponds exactly
to the dipole swing in fig.~\ref{fig:swing},
and consequently it has exactly the same effect as the gluon 
exchange process discussed above. The only difference we may expect is that 
the effect due 
to multipoles would be instantaneous, while the gluon exchange process 
ought to be proportional to the rapidity interval, $\Delta y$, 
in the evolution. 

We have not (yet) found a weight for the dipole swing which makes the
scattering process explicitely frame independent. We note, however, that
the dipole splitting vertex in \eqref{eq:dipkernel} gives a total
weight for a specific dipole chain given by the product of factors
$1/\mathbf{r}_i^2$ for all dipoles still present in the cascade.
For dipoles which have already decayed (those denoted by dashed
lines in figs.~\ref{fig:multcoll} and \ref{fig:dipswing}) 
the factor $\mathbf{r}_i^{-2}$
is exactly compensated by the factor $\mathbf{r}_i^2$ in the numerator
of the splitting vertex factor. It may therefore seem natural that
the swing $(\pmb{x}_1,\pmb{y}_1)+(\pmb{x}_2,\pmb{y}_2) \rightarrow 
(\pmb{x}_1,\pmb{y}_2)+(\pmb{x}_2,\pmb{y}_1)$ has 
 a weight proportional to 
$(\pmb{x}_1-\pmb{y}_1)^2(\pmb{x}_2-\pmb{y}_2)^2/(\pmb{x}_1-\pmb{y}_2)^2(\pmb{x}_2-\pmb{y}_1)^2$.
Such a weight would preserve the
feature that any dipole system has a weight proportional to
$\prod \mathbf{r}_i^{-2}$. At the same time it also favours dipoles
formed by close charge--anti-charge pairs in colour quadrupoles,
in accordance with the discussion above. 

\subsubsection{Implementation of the Dipole Swing}
\label{sec:implswing}
When we include the dipole
swing in the MC implementation we will therefore use a weight proportional
to $(\pmb{x}_1-\pmb{y}_1)^2(\pmb{x}_2-\pmb{y}_2)^2/(\pmb{x}_1-\pmb{y}_2)^2(\pmb{x}_2-\pmb{y}_1)^2$, 
and the 
normalization should be adjusted so that the final result is approximately
frame independent. This would show that there is the same probability
to have a colour loop within the cascade evolution as formed by multiple 
sub-collisions. Even if this is achieved it is, however, not possible
to tell whether the dominant mechanism behind the swing is
due to gluon exchange or to colour multipoles.

In the MC the probability for a swing 
$(\pmb{x}_1,\pmb{y}_1)+(\pmb{x}_2,\pmb{y}_2) \rightarrow (\pmb{x}_1,\pmb{y}_2)+(\pmb{x}_2,\pmb{y}_1)$
is formulated as if the main mechanism is colour multipoles. There is a probability 
$1/(N_c^2 - 1)$ that two dipoles have the same colour. Therefore 
we assume that two dipoles are allowed to swing with this probability.
If they are allowed to swing, they do so during an evolution step $\Delta y$
with a probability given by
\begin{equation}
\Delta P = \lambda 
\frac{(\pmb{x}_1-\pmb{y}_1)^2(\pmb{x}_2-\pmb{y}_2)^2}{(\pmb{x}_1-\pmb{y}_2)^2(\pmb{x}_2-\pmb{y}_1)^2}
\Delta y
\label{eq:probswing}
\end{equation}
If the normalization factor $\lambda$ is large, the dipoles with
the same colour will swing
rapidly and adjust themselves to weights $\propto \prod \mathbf{r}_i^{-2}$
before the next dipole splitting. The applications presented
in section \ref{sec:results} are obtained using such a large $\lambda$-value,
which implies that the swing can be regarded as instantaneous,
as expected if colour multipoles is the dominant mechanism.
This also means that the effective normalization is given by the number of 
different dipole colours $N_c^2 - 1 = 8$ and not by $\lambda$. 
We will see that this recipe does
indeed give an almost frame independent result. This does, however,
not imply that we also can conclude that colour multipoles is the dominant 
mechanism. It is actually possible to get approximately the same result 
allowing swings
between more dipoles but with a smaller $\lambda$-value. In this 
case the swing does not occur instantaneously, but with a probability
proportional to the step $\Delta y$ in rapidity, as expected if the 
dominant mechanism is gluon exchange. It is therefore not possible
to tell which mechanism is most important.

A more detailed description of how the swing is implemented in the 
simulation program is given in section \ref{sec:MC}.

\subsection{Comparison With Other Formalisms}
 
There has been quite some effort to interpret pomeron mergings as 
dipole mergings, i.e. interpreting \eqref{eq:pomloopev}
in terms of a system of dipoles which can either split (a $1\rightarrow 2$
dipole vertex) or merge (a $2\rightarrow 1$ vertex).
While it is obvious that dipole mergings generate pomeron mergings, 
the opposite of this statement is not necessarily true. In fact
there are also other dipole processes which generate pomeron
mergings. In a formalism where the cascade is treated as a set
of uncorrelated dipoles a $2\rightarrow n$ dipole vertex, 
with $n \geq 2$, also leads to pomeron mergings. This follows because there is 
always the possibility that only one of the new dipoles 
interacts with the target, while the rest are spectators. Of course 
such a transition leads to all possible $2 \rightarrow m$ 
($m = 1, \dots,n$) pomeron transitions. 

This argument 
can be illustrated by the following evolution equations. 
We denote the dipoles by the letters $a,b$ and so on, and 
the $S$-matrix by $\mathcal{S}(a)$, for one dipole $a$, $\mathcal{S}(ab)$
for two dipoles $a$ and $b$. The scattering amplitude
is given by $T=1-\mathcal{S}$ and with $T(ab)$ we mean 
$\langle T(a)T(b)\rangle$. Assume now that there exist different
vertices for different dipole transitions; $\beta (ab\vert c)$
for the merging of $a$ and $b$ into $c$, $\Gamma (ab\vert cd)$
for the transition $a+b \rightarrow c+d$ and so on. We then have the 
following evolution equations 
\begin{eqnarray}
\partial_Y \mathcal{S}(ab)_\beta&=&\int_{c} \beta (ab\vert c) \{ -\mathcal{S}(ab)
+\mathcal{S}(c) \}  \nonumber \\
\partial_Y \mathcal{S}(ab)_\Gamma&=&\int_{cd} \Gamma (ab\vert cd) \{ -\mathcal{S}(ab)
+\mathcal{S}(cd) \}.
\end{eqnarray}
The negative contribution on the right hand side comes from the 
fact that the system can remain as it is, with a survival probability given by
$(1-\int \beta)$ or $(1-\int \Gamma)$. Alternatively the system can evolve, 
with a probability density given by $\beta$ or $\Gamma$, which 
corresponds to the positive contribution on the right hand side. 
Thus the evolution of the $S$-matrix has a clear probabilistic 
interpretation. One can now rewrite these equations for $T$, using 
the relation $T=1-\mathcal{S}$. We thus get
\begin{eqnarray}
\partial_Y T(ab)_\beta &=& \int_{c} \beta (ab\vert c) \{ -T(c)
+T(a) + T(b) - T(ab) \}  \nonumber \\
\partial_Y T(ab)_\Gamma &=& \int_{cd} \Gamma (ab\vert cd) \{ -T(c)
-T(d)+T(a)+T(b)-T(ab)+T(cd)\}.
\label{eq:dyT}
\end{eqnarray}
Here we can see that both equations contain both 
pomeron mergings and also $2\rightarrow 2$ pomeron transitions.
It is straightforward to write the equations also for more general 
vertices. Indeed, for the general $2\rightarrow n$ transition, with 
the vertex $\Gamma (ab\vert c_1c_2\dots c_n)$, we get the 
following evolution equations
\begin{eqnarray}
\partial_Y \mathcal{S}(ab)_\Gamma&=&\int_{c_i} \Gamma (ab\vert c_1c_2\dots c_n)
 \{ -\mathcal{S}(ab)+\mathcal{S}(c_1c_2\dots c_n) \} \nonumber \\
\partial_Y T(ab)_\Gamma &=& \int_{c_i} \Gamma (ab\vert c_1c_2\dots c_n)
 \{T(a)+T(b)-T(ab)+ \nonumber \\ 
&+& \sum_{k=1}^n(-1)^k\sum_{i_k>i_{k-1}>\dots >i_1}^n
T(c_{i_1}c_{i_2}\dots c_{i_k}) \}.
\label{eq:dyT2}
\end{eqnarray}
The interpretation of the equations for the amplitude $T$ in terms 
of pomeron transitions can however be misleading, especially if 
a single dipole is allowed to couple to several pomerons. To find 
the equation for $T$, it is always safer to start with the 
corresponding equation for $\mathcal{S}$ and then use the relation
$T=1-\mathcal{S}$ to deduce the equation for $T$, just as we have 
done above. The equation for $\mathcal{S}$ is determined by the 
structure of the corresponding dipole vertex and has a simple 
interpretation as described above. 

Note that so far we have 
not asked whether equation \eqref{eq:pomloopev} 
can be rewritten in a similar way. As mentioned above, this has 
been attempted by trying to write it
with a contribution of the form $\partial_Y T(ab)_\beta$ in (\ref{eq:dyT}). 
However, it was shown in \cite{Iancu:2005dx} that this approach
has problems. Formally it is possible, 
but the problem is that the would-be dipole merging vertex (in this 
case the $\beta$-vertex above) has no fixed sign as is 
required in a proper probabilistic formalism.

We now want to demonstrate that the dipole swing discussed in the
previous subsection can reproduce not only pomeron merging and
loop formation, but also the more complicated transitions described
in \eqref{eq:dyT2}. Since 
the dipole model is just an effective picture it is likely  
that a more complete treatment will involve more general vertices, 
generating transitions involving an arbitrary number of pomerons. 
As discussed in section \ref{sec:implswing}, the 
weight for this process involves a phenomenological normalization parameter
$\lambda$ which determines the strength of the process, i.e. how 
fast this process happens in rapidity. In the applications
presented in section \ref{sec:results} the value of $\lambda$ is such that 
the recouplings saturate in the sense that the recouplings lead to an 
equilibrium given by the weights proportional to $\prod \mathbf{r}_i^{-2}$.
This means that effectively these recouplings happen instantaneously. After 
each gluon emission the cascade will evolve through recouplings 
back and forth until it settles in a preferred configuration (the weight in 
\eqref{eq:probswing} here favours configurations where the dipoles are 
as small as possible) after which
there is a new gluon emission (dipole splitting). 

Assume that, at some 
rapidity, we have $N$ partons, located at positions 
$(\pmb{x}_0,\pmb{x}_1,\dots ,\pmb{x}_{N-1})$ where $\pmb{x}_0$ and $\pmb{x}_{N-1}$ 
are the positions of the initial quark and antiquark respectively. 
Assume now that a gluon $\pmb{z}$ is emitted from some dipole
$(\pmb{x}_i,\pmb{x}_j)$ which means that this dipole is replaced 
by $(\pmb{x}_i,\pmb{z})$ and $(\pmb{z},\pmb{x}_j)$. After this there 
will be a series of recouplings which transform the cascade into some
new configuration involving $N$ dipoles. From our discussion above, 
it follows that these recouplings will most likely involve the new 
dipoles which were produced after the emission of $\pmb{z}$. This is
so because the cascade, prior to the emission of $\pmb{z}$, already 
has settled in a preferred configuration and, apart from the 
replacement of $(\pmb{x}_i,\pmb{x}_j)$ with $(\pmb{x}_i,\pmb{z})$ and 
$(\pmb{z},\pmb{x}_j)$, it keeps the same configuration after $\pmb{z}$ is emitted. 
Therefore it is not so likely to have further recouplings among the other dipoles 
(if not, these would most likely have happened before the emission of $\pmb{z}$).
There will thus be a series of recouplings involving newly produced dipoles 
until the cascade once again settles in some preferred configuration, 
after which there will be a new emission. 

The discussion above suggest that we may view the evolution as effectively
being driven by vertices involving $k\rightarrow k+1$ dipole transitions, where
$k$ is determined by how many swings we have between the emissions. 
For a cascade evolving through such a general vertex we can 
write the evolution equations for $\mathcal{S}$ and $T$ just as we did 
in above. Once again we denote the dipoles with letters 
$a_i$ and the corresponding $S$-matrices with $\mathcal{S}(a_i)$ for 
a single dipole, $\mathcal{S}(a_ia_j)$ for two dipoles and so on. For 
a vertex $\Gamma_k (a_1\dots a_k| b_1\dots b_{k+1})$, giving rise to the
transition $a_1+\dots +a_k\rightarrow b_1+\dots +b_{k+1}$ we then get 
the following evolution equations 
\begin{eqnarray}
\partial_Y \mathcal{S}(a_1\dots a_k)_{\Gamma_k}&=&\int_{b_i}\Gamma_k (a_1\dots a_k|b_1\dots b_{k+1})
\{ -\mathcal{S}(a_1\dots a_k)+\mathcal{S}(b_1\dots b_{k+1}) \} \nonumber \\
\partial_Y T(a_1\dots a_k)_{\Gamma_k}&=&\int_{b_i}\Gamma_k (a_1\dots a_k|b_1\dots b_{k+1})
\{\sum_{m=1}^k(-1)^{k+m-1}\sum_{i_m>\dots >i_1}^kT(a_{i_1}\dots a_{i_m}) + \nonumber \\
&&+ \sum_{m=1}^{k+1}(-1)^{k+m}\sum_{i_m>\dots >i_1}^{k+1}T(b_{i_1}\dots b_{i_m})\}.
\label{eq:kdipvert}
\end{eqnarray}
Here one sees that $T^k$ is coupled to $T^m$ with $m=1, \dots, k+1$ which 
means that there are pomeron mergings of the type $k\rightarrow i$, $i=1, \dots, k-1$. 
We also note that similar type of equations involving some more 
general vertices have recently been presented in \cite{Kozlov:2006cg,Blaizot:2006wp} although 
the structure of the vertices are different. If we consider the 
process in zero transverse dimensions, which defines the toy model 
first presented in \cite{Mueller:1994gb} (see also \cite{Kovner:2005aq}), 
then for all dipoles $a_i$ one replaces 
$T(a_i)$ by some $t$ (and $T(a_1\dots a_k)$ by $t^k$) which is the amplitude 
in the toy model. One can then 
see that the equations for $t^k$ presented in \cite{Blaizot:2006wp} (equations 
(2.19) to (2.21)) can be understood in terms of the general transitions in 
\eqref{eq:kdipvert}. 
 
\section{The Initial States in the Cascade Evolutions}
\label{sec:proton}

In the introduction we argued that results from the Tevatron supports
a picture where high energy $pp$ interactions are dominated by perturbative 
parton--parton sub-collisions. This encourages us to interpret $\gamma^*p$ and 
$pp$ collisions as a result of perturbative dipole interactions.
We then also need an
initial dipole state for a virtual photon and for a proton. 

\subsection{The Virtual Photon}

The coupling of a virtual photon to a quark--antiquark pair is well known.
In leading order and for the case of massless quarks the longitudinal and 
transverse wave functions, denoted $\psi_L$ and $\psi_T$ respectively,
are given by
\begin{eqnarray}
\vert \psi_L(z,r)\vert^2&=&\frac{6\alpha_{em}}{\pi^2}\sum_q 
e_q^2Q^2z^2(1-z)^2K_0^2(\sqrt{z(1-z)}Qr) \nonumber \\
\vert \psi_T(z,r)\vert^2&=&\frac{3\alpha_{em}}{2\pi^2}
\sum_qe_q^2[z^2+(1-z)^2]z(1-z)Q^2K_1^2(\sqrt{z(1-z)}Qr).
\label{eq:psigamma}
\end{eqnarray}
Here $z$ ($1-z$) is the longitudinal momentum fraction of the quark (antiquark) 
and $\pmb{r}$ is the transverse separation between them. $Q^2$ denotes
the virtuality of the virtual photon and $K_0$ and $K_1$ are  
modified Bessel functions. The sum in \eqref{eq:psigamma} runs over 
all active quark flavours, and in our case we consider 3 massless 
flavours.

\subsection{The Initial Proton}

The initial state for the proton can naturally not be determined by
pertubation theory, but has to be described by some model assumption,
which can obviously never fully describe all features of the proton.
We have tried a couple of different approaches based on the assumption
that a proton at rest mainly consists of its three valence quarks. It
is natural to think of these three quarks as the endpoints of dipoles
and, assuming that the quarks all have different colours, this would
give three different dipoles. One would not expect these dipoles to be
completely independent, and indeed it was argued in
\cite{Praszalowicz:1997nf} that they would be non-trivially
correlated. We have tried two different approaches with varying degree
of correlation\footnote{In the original dipole formulation, all
  dipoles are independent and correlations can only be introduced
  through their relative placement in impact-parameter space. However,
  when introducing explicit energy conservation, neighboring dipoles
  will affect each other, thus introducing an additional
  correlation.}: completely uncorrelated dipoles or a ``triangle''
configuration. In the latter case each quark is connected by two
dipoles to the other two quarks, ie.\ assuming that eg.\ a red quark
behaves essentially as an anti-blue--anti-green colour compound to
form dipoles with the other two (blue and green) quarks.

One could argue that a more natural choice would be to use a
``Mercedes'' star configuration, where all three dipoles are joined in
a junction. However, a picture where the junction does not carry energy 
and momentum is then difficult to reconcile with the dipole formalism, 
and we have in the following
settled for the triangle configuration with the dipole sizes 
distributed as Gaussians with an average size $\sim 3.1$~GeV$^{-1}$, 
which is determined by a fit to $pp$ data. 

\section{Monte Carlo Simulation}
\label{sec:MC}

Our calculations are performed with a simple simulation program
written in C++, where there are gluons connected by dipoles and vice
versa. A dipole state is described by a set of partons, each of which
has a specified position in the transverse plane and a rapidity, $y$
(which is the true rapidity and not $\log(1/x)$).
In addition, each gluon is assigned a transverse momentum when it is
emitted, corresponding to the inverse of the transverse size of the
smallest dipole to which it is connected. Thus, if we have a splitting
as in \eqref{eq:dipkernel} we assign
\begin{equation}
  p_\perp=\frac{2}{\min(|\pmb{x}-\pmb{z}|,|\pmb{z}-\pmb{y}|)}.
  \label{eq:pt}  
\end{equation}
In this way we can implement ordering in $p_\pm=p_\perp e^{\pm y}$
separately. The transverse momentum of each of the emitting partons
will be set to 2 times the inverse transverse distance to the emitted
gluon if this is larger than the previously assigned $p_\perp$. In
addition the rapidities of the emitting partons are changed so that
the total positive light-cone momentum component is conserved in each
emission.  These recoils are distributed so that the emitting parton
at $\pmb{x}$ contributes a fraction
$|\pmb{z}-\pmb{y}|/(|\pmb{x}-\pmb{z}|+|\pmb{z}-\pmb{y}|)$ to the $p_+$
of the emitted gluon. The assignment of the $p_\perp$ for the gluons
as given above, and the conservation of the $p_+$ component,
automatically gives a cutoff for small dipoles. Therefore we do not
need to explicitely introduce a cutoff $\rho$, as described in section
\ref{sec:Muellerdip}, but we rather obtain a \emph{dynamical cutoff},
$\rho(\Delta y)$, which is large for small steps in rapidity, $\Delta
y$, but gets smaller for larger $\Delta y$.

In each step an emission is generated for each of the dipoles in a
state according to \eqref{eq:dipkernel} and the corresponding Sudakov
form factor, allowing $\alpha_s$ to run according to the one-loop
expression with the scale set to the $p_\perp$ of the emitted
gluon\footnote{To avoid divergencies, $\alpha_s$ is frozen below
  the scale $p_\perp=2/r_{\max}$.}.  The dipole which has generated
the smallest step in rapidity is then allowed to radiate and is
replaced by two new dipoles, and the procedure is reiterated until no
dipole has generated a rapidity above (or below, depending of the
direction of evolution) a minimum (or maximum) rapidity. Finally a
cross section can be calculated by letting the dipoles from the target
and the projectile, both properly evolved, collide at a random impact
parameter according to \eqref{eq:dipamp} and \eqref{eq:sigmaonioni}.

In the dipole model it is possible to create arbitrarily large
dipoles. Even if the $p_-$ ordering in our formalism sets a limit for
how large a dipole can be, just as the $p_+$ ordering sets a limit for
how small a dipole can be, there is no mechanism suppressing the
formation of large dipoles. On the contrary they are enhanced by the
running of $\alpha_s$. Obviously confinement must suppress the
formation of larger dipoles and we therefore choose a parameter
$r_{max}$ such that each emission is suppressed with a Gaussian of
average size $\sim r_{max}$.  This means that each emission, giving a
new dipole of size $r$, is allowed with a probability
exp$(-r^2/r_{max}^2)$. We choose $r_{max}$ to have the same value as
the average size of the initial dipoles in the proton, i.e.
$r_{max}=3.1$GeV$^{-1}$, as this corresponds to the nonperturbative
input for the proton.

When implementing the dipole swing mechanism we followed the strategy
introduced in the Ariadne program
\cite{Lonnblad:1995wk,Lonnblad:1992tz} where each dipole is randomly
assigned a colour index in the range $[1,N_c^2-1]$ in such a way that
no neighboring dipoles have the same index.\footnote{Naively one might
  expect there only to be three differently coloured dipoles, but the
  probability that two arbitraty dipoles are allowed to recouple is
  related to the number of different gluons rather than to the number
  of colours which is why we have $N_c^2-1$ different dipole indices.}
In each step any pair of dipoles with the same index is allowed to
generate a fictitious rapidity for a recoupling according to
\eqref{eq:probswing} modified with an appropriate Sudakov form factor.
These generated recouplings are then allowed to compete with the
generated emissions, so that in each step the process giving the
smallest step in rapidity is performed. Because of the way colour
indices are assigned we can ensure that no unphysical dipole chains
(eg.\ with colour-singlet gluons) can occur.

\FIGURE[t]{
  \includegraphics[angle=270, scale=0.72]{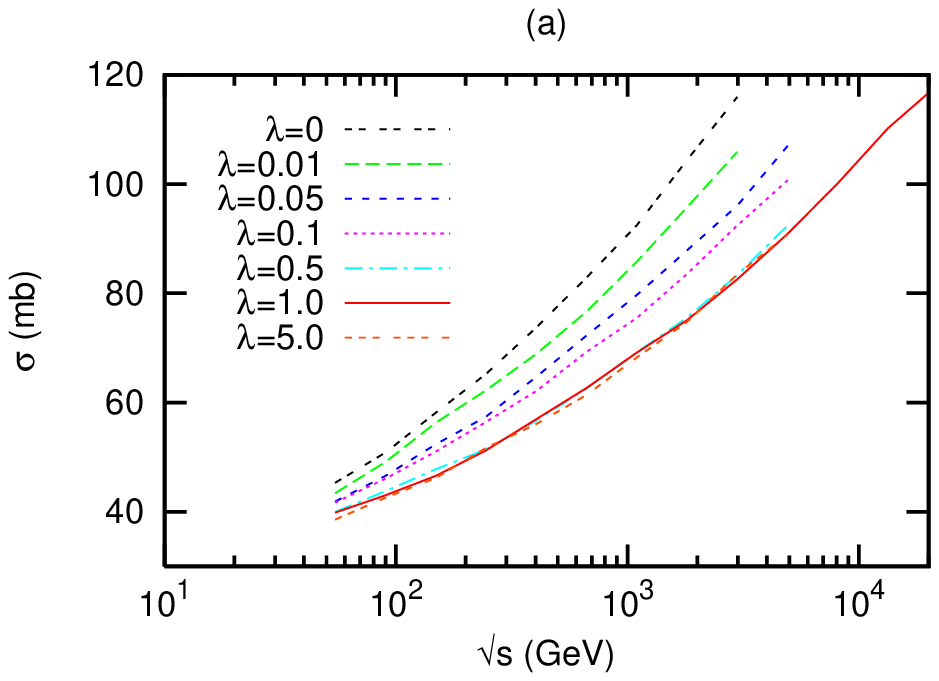}
  \includegraphics[angle=270, scale=0.72]{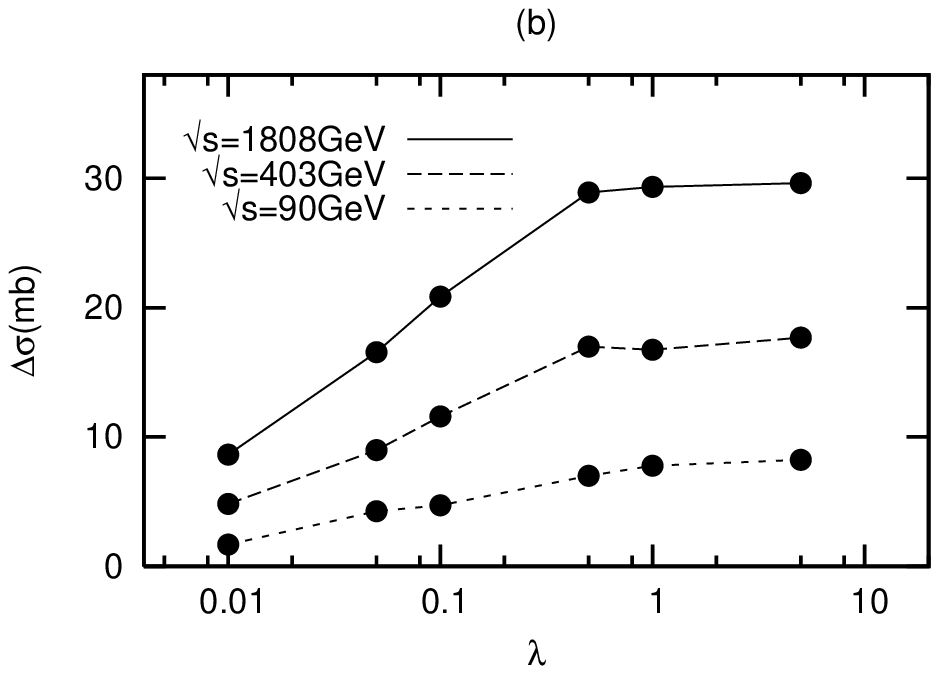}
  \caption{\label{fig:lambda} The $pp$ cross section for 
    different values of the coefficient $\lambda$. It is seen in (a)
    that $\sigma$ does not change for $\lambda \gtrsim 0.5$. This can
    also be seen from (b) which shows the difference in the cross
    section, $\Delta \sigma (\lambda) \equiv \sigma (\lambda =0)-\sigma
    (\lambda)$,as a function of $\lambda$, and for different $s$.}  }

Clearly if $\lambda$ in \eqref{eq:probswing} is very large, the
evolution is swamped by recouplings back and forth, making the
simulation very inefficient. In this way we also see that the effect
of the recouplings must saturate at large enough $\lambda$. By chance
it turns out that a value of $\lambda=1$ is just large enough for
saturation, see figure \ref{fig:lambda}.

It can be shown that two dipoles recoupling back and forth in this
saturated way and colliding with a single dipole according to
\eqref{eq:dipamp} corresponds exactly to a quadrupole--dipole
scattering. Also higher multipole--multipole scatterings are in this
way well approximated.

As discussed above these recouplings in some sense also give rise to
pomeron mergings, as configurations where one dipole is very small is
favored and this dipole has a smaller probability to interact with
the target. In our program it is also possible to include explicit
mergings of neighboring dipoles, a process which is necessary for the
study of exclusive final states and will be studied in a future
publication. It should be noted that the combined process of first
splitting a dipole into two, then recoupling with a third dipole and
finally merging one of them again, corresponds to the recoupling of
two dipoles with different colour indices by the exchange of a gluon.

\section{Results}
\label{sec:results}

In this section we compare results obtained from our model with 
experimental data from DIS at HERA and $pp$ collisions at the Tevatron. 
As we yet have not a full control over 
the virtual dipoles we only study the total cross sections.

\subsection{DIS}
 
In \cite{Avsar:2005iz} we obtained a reasonable qualitative agreement with 
HERA data for the $F_2$ structure function and the effective slope 
$\lambda_{\mathrm{eff}}(Q^2) = d (\log F_2 )/d (\log 1/x)$. Having now
improved our model further, we will see that we also obtain a 
good quantitative agreement with the data. 

The $\gamma^*p$ total cross section is given by 
the sum of the two wave functions in \eqref{eq:psigamma},
convoluted with the dipole--proton cross section, $\sigma(z,\pmb{r})$, 
\begin{eqnarray}
\sigma_{\gamma^*p}^{tot}=\int d^2\pmb{r}\int_0^1 dz\{\vert \psi_L(z,r)\vert^2
+\vert \psi_T(z,r)\vert^2\} \sigma(z,\pmb{r}).
\end{eqnarray}
Here the dipole--proton cross section $\sigma(z,\pmb{r})$ 
is obtained from equations \eqref{eq:sigmaonioni} and 
by \eqref{eq:dipamp}, when the initial proton state in section \ref{sec:proton}
is evolved as described in section \ref{sec:MC}.

\FIGURE[t]{
  \includegraphics[angle=270, scale=0.645]{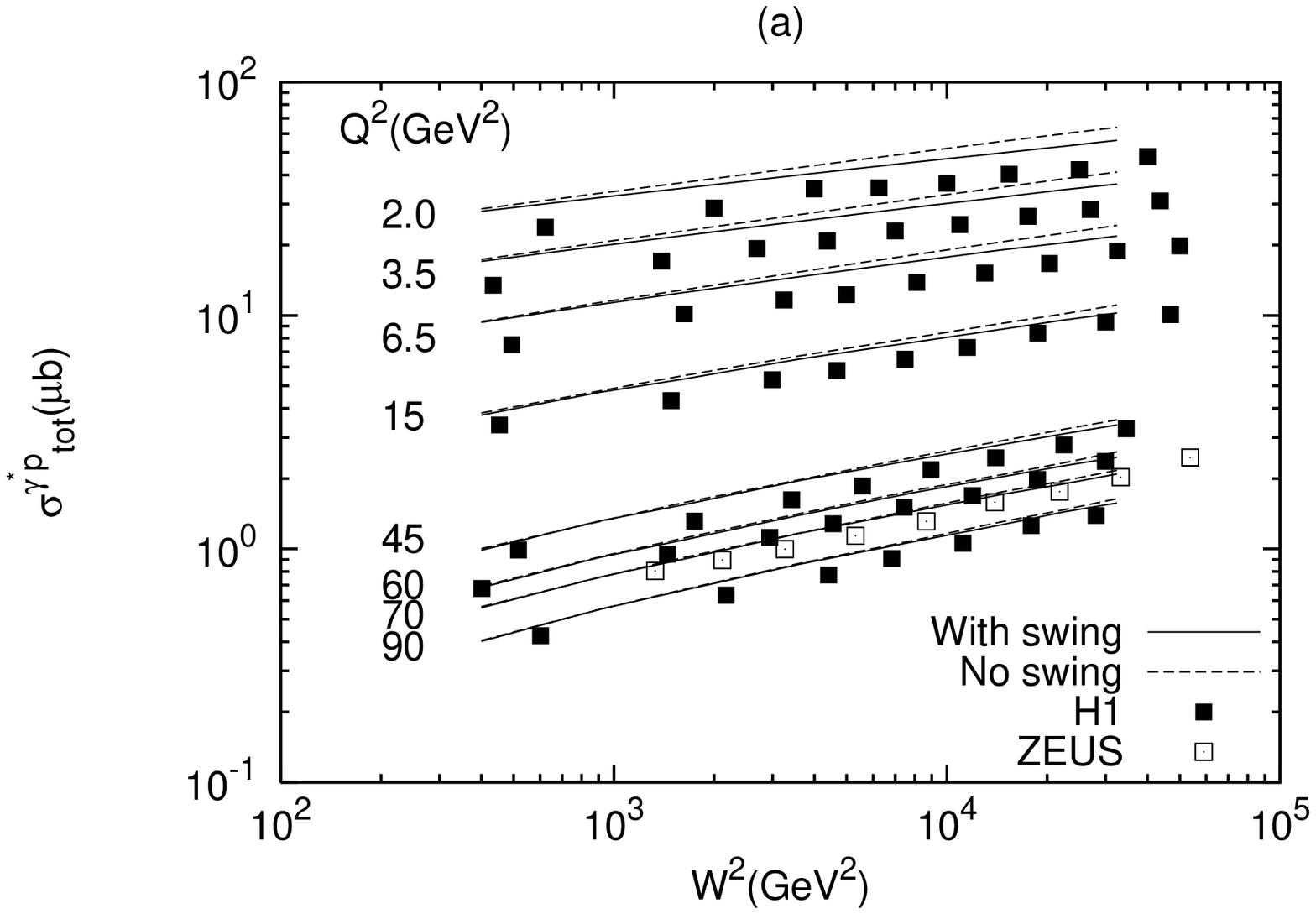}\\
  \includegraphics[angle=270, scale=0.6]{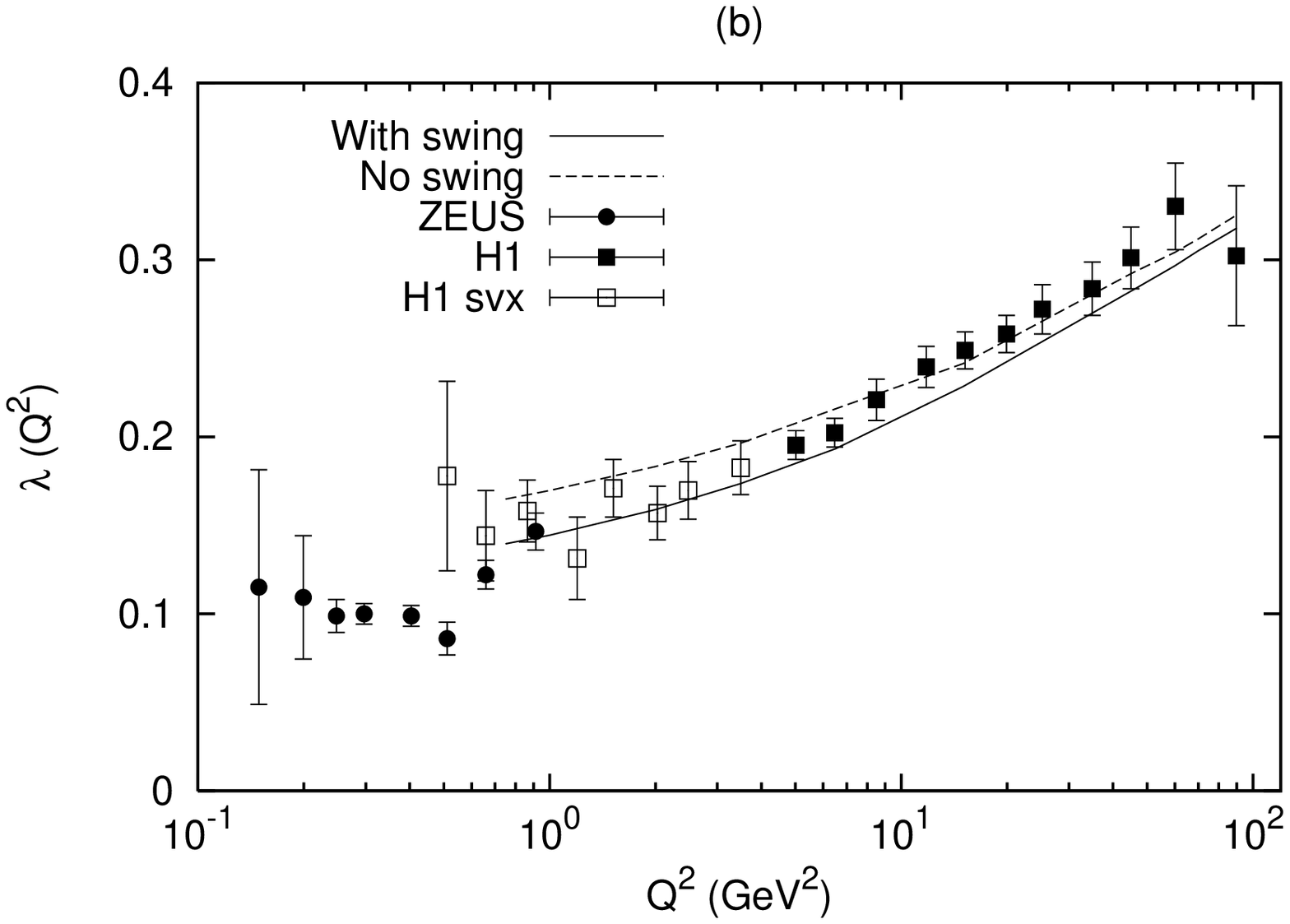}
  \caption{\label{fig:gptot}(a) The $\gamma^*p$ total cross section shown for 
    different $Q^2$. The solid lines include the dipole swing mechanism
    while the dashed lines do not. $W$
    denotes the cms energy. (b) The effective slope measured at
    different $Q^2$. The lines are the same as in (a). Filled
    circles are data from ZEUS\cite{Breitweg:2000yn} while
    filled\cite{Adloff:2000qk} and open\cite{DIS04Petrukhin} squares
    are data from H1.}  }

In figure \ref{fig:gptot}a we show the results for the $\gamma^*p$
total cross section. As we can see the results are in quite good
agreement with data except for the fact that the normalization is
around 10--15\% too high for $Q^2\lesssim$ 15 GeV$^2$ while it is
around 5--10\% too high for $Q^2\gtrsim$ 45 GeV$^2$.  Data points are
taken from \cite{Adloff:2000qk,Breitweg:2000yn} and we also see that
the effects of dipole swings are quite small, mainly visible for
$Q^2\lesssim$ 15 GeV$^2$.

As seen from figure \ref{fig:gptot}a, our results seems to have the
correct $W$ dependence. This can be seen more clearly from figure
\ref{fig:gptot}b, where we show the results for the logarithmic
slope $\lambda_{\mathrm{eff}}=d (\log \sigma)/d (\log 1/x)$.  We see
that there is a good agreement with data for all points lying in the
interval 1GeV$^2 \lesssim Q^2 \lesssim$100 GeV$^2$. Here the slope is
determined within the same energy interval from which the experimental
points, taken from \cite{Breitweg:2000yn,Adloff:2000qk,DIS04Petrukhin},
are determined.

\subsection{Proton--Proton Collisions}

\FIGURE[t]{
  \includegraphics[angle=270, scale=0.7]{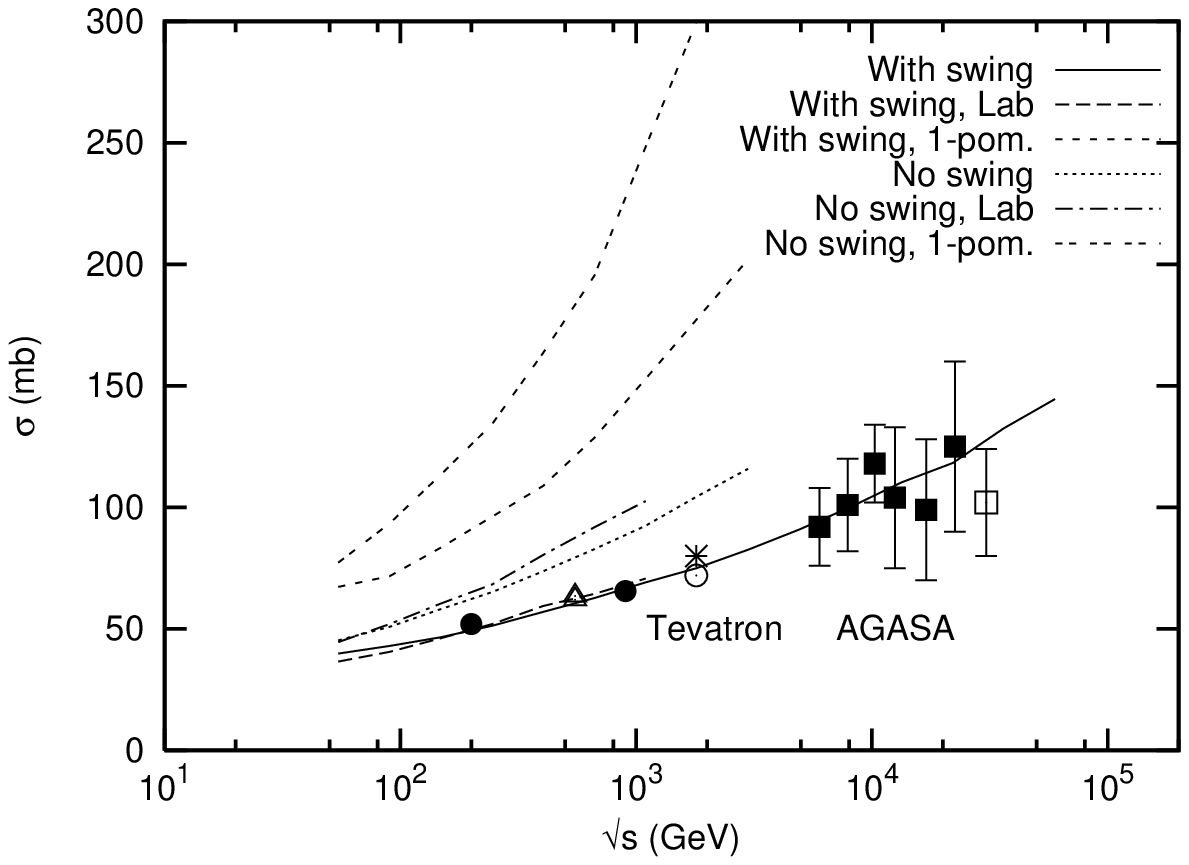}
  \caption{\label{fig:pptotal}The total cross section for $pp$ scattering as 
    a function of the cms energy $\sqrt{s}$. Here results are shown 
    for evolution with and without the dipole swing mechanism. The results 
    for the one pomeron contribution are also shown. Also shown are
    the results obtained in the ``lab'' frame where one of the protons is 
    almost at rest.}
}

\FIGURE[t]{
  \includegraphics[angle=270, scale=0.45]{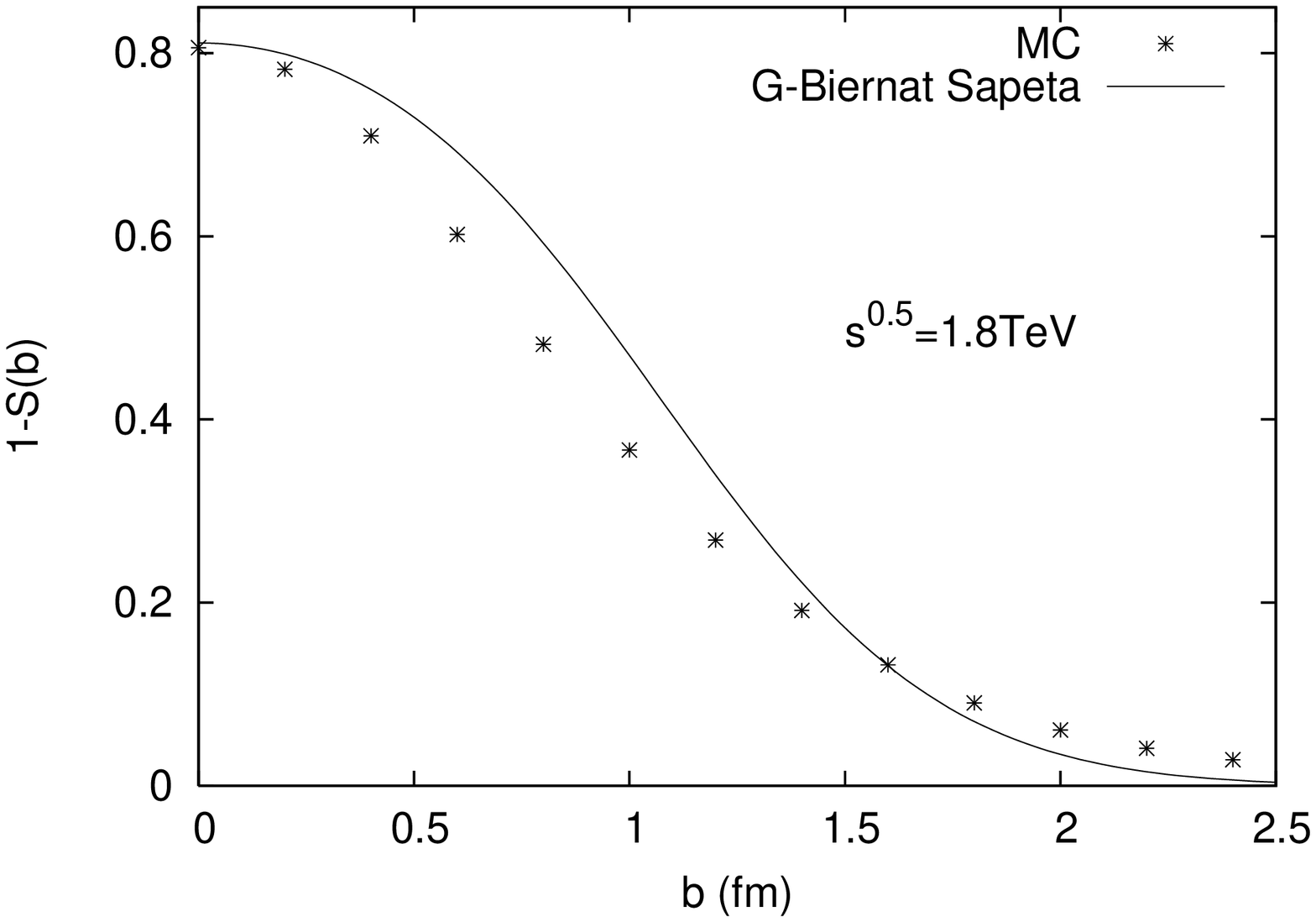}
  \caption{\label{fig:ppbdep}The $pp$ impact parameter profile, 
    $(1/2)d\sigma_{tot}/d^2b$. The points are results obtained 
    from our Monte Carlo while the line is a two parameter Gaussian fit from 
    ref.~\cite{Sapeta:2005ba}.}
}

The results obtained for the $pp$ total 
cross section are presented in fig.~\ref{fig:pptotal}. Here we see 
that the dipole swing have a
rather large effect, as expected. In the figure we also 
show the results for the one pomeron cross sections, and one can 
see the large effects of unitarisation. Our main results are calculated 
in the center of mass frame, where the colliding protons share the
energy equally, but in the figure we also show results obtained 
in the ``lab'' frame, where one of the protons carries almost all 
avaliable energy, while the other one is essentially at rest. Due to the 
fact that the Monte Carlo simulation becomes very inefficient in 
such a frame (since the energetic proton has to be boosted to 
quite high rapidities) we have evaluated $\sigma_{tot}$ only 
up to $\sqrt{s}\sim1$~TeV. Although the result is not exactly
frame independent we see that the frame dependence is reduced,
and now very small.  

The final result of this paper concerns the impact parameter 
dependence of the $pp$ total cross section. The result is shown
in figure \ref{fig:ppbdep}. Here we have plotted $(1/2)d\sigma_{tot}/d^2b$ 
as a function of $b$ and for $\sqrt{s} = 1.8$TeV. The result is 
compared to a two parameter Gaussian fit from ref.~\cite{Sapeta:2005ba}.
There is a quite good agreement between the results.
Our profile is more flat and has a somewhat longer tail, but a Gaussian
fit to our result would be very similar to the fit in 
ref.~\cite{Sapeta:2005ba}.

\section{Conclusions}
\label{sec:conc}

The QCD description of high-energy scattering is clearly a complicated
subject. Although a qualitative description of inclusive cross
sections for $\gamma^*p$ and $pp$ scatterings can be obtained in eg.\ 
the B--JIMWLK formalism, it is difficult to give quantitative
predictions, and even more difficult to describe exclusive properties
of the final states.

In this paper we have described a model of interacting colour dipoles
based on a limited set of fairly simple ingredients:
\begin{itemize}\itemsep 0mm
\item The description of the initial-state virtual photon and proton
  as a (set of) dipole(s).
\item Simple dipole splittings by gluon emissions according to the
  Mueller formalism.
\item Energy--momentum conservation in each splitting, which gives a
  dynamic cutoff for small dipole sizes and introduces non-trivial
  correlations between neighboring dipoles.
\item A mechanism for dipole reconnections (or ``swing'')
  corresponding to the introduction of pomeron loops in the evolution.
\item A simple dipole--dipole scattering cross section, exponentiated
  to include multiple scattering and saturation effects.
\end{itemize}

Although each ingredient is fairly simple, it is
prohibitively difficult to include them all in analytic calculations.
However, implementing them in a Monte Carlo simulation program we are
able to reproduce to a satisfactory degree both the $\gamma^*p$ cross
sections as measured at HERA as well as the total $pp$ cross section
all the way from ISR energies to the Tevatron and beyond. It should be
pointed out that this is achieved with effectively only two tuneable
parameters, the basic QCD scale $\Lambda_{\mbox{\scriptsize QCD}}$ and
$r_{\max}$ giving the non-perturbative cutoff for large dipoles. There
are two additional parameters involved in the dipole swing. One
is $\lambda$ and we have shown that as long as it is large enough to
saturate the recouplings, the results are completely insensitive to
this parameter. The other parameter is the number of colour indices,
$N_c^2-1$, which we have fixed to 8, but could in principle be
considered a free parameter to simulate the effect of recouplings
between differently coloured dipoles by gluon exchange.

The resulting description is quite insensitive to the Lorentz frame
chosen to perform the simulations, which shows that we have a
consistent treatment of pomeron loops in the evolution and in the
scattering of evolved dipole states. 

Recently there has been a lot of activity in the subject of 
high energy evolution in QCD. Different, and very interesting, models
have been proposed which are often based on analytical calculations in some
toy limit or at asymptotically high energies. We here want to emphasize
the importance of a working Monte Carlo simulation in order to confront 
QCD with real data at realistic energies.
 
We have also seen that the way the dipole swing is implemented 
in our model makes it possible to view the evolution as effectively
being driven by more general vertices which give rise to more 
general pomeron transitions during the evolution. Some ideas with 
such vertices have recently been presented in \cite{Kozlov:2006cg, Blaizot:2006wp}. 
There is still more work to do with regard to explicit frame 
independence in the dipole model, and it is our intention to further
study this problem in future investigations.   

Another advantage of our model is that it should be possible to also
simulate exclusive properties of the final states. Confronting these
with experimental observables will allow us to gain further insight
into the QCD evolution at high energies. We will return to this problem in a
future publication.

\bibliographystyle{utcaps}
\bibliography{/home/shakespeare/people/leif/personal/lib/tex/bib/references,refs}

\providecommand{\href}[2]{#2}\begingroup\raggedright\begin{thebibliography}{10%
}\itemsep 0mm

\bibitem{Andersen:2006sp}
J.~R. Andersen {\em Phys. Lett.} {\bf B639} (2006) 290--293,
\href{http://www.arXiv.org/abs/hep-ph/0602182}{{\tt hep-ph/0602182}}.

\bibitem{Mueller:1981ex}
A.~H. Mueller {\em Phys. Lett.} {\bf B104} (1981)
161--164.

\bibitem{Ermolaev:1981cm}
B.~I. Ermolaev and V.~S. Fadin {\em JETP Lett.} {\bf 33} (1981)
269--272.

\bibitem{Bassetto:1982ma}
A.~Bassetto, M.~Ciafaloni, G.~Marchesini, and A.~H. Mueller {\em Nucl. Phys.}
  {\bf B207} (1982)
189.

\bibitem{Marchesini:1983bm}
G.~Marchesini and B.~R. Webber {\em Nucl. Phys.} {\bf B238} (1984)
1.

\bibitem{Gustafson:1986db}
G.~Gustafson {\em Phys.~Lett.} {\bf B175} (1986)
453.

\bibitem{Gustafson:1988rq}
G.~Gustafson and U.~Pettersson {\em Nucl.~Phys.} {\bf B306} (1988)
746.

\bibitem{Abbiendi:2003ri}
{\bf OPAL} Collaboration, G.~Abbiendi {\em et al.} {\em Eur. Phys. J.} {\bf
  C35} (2004) 293--312,
\href{http://www.arXiv.org/abs/hep-ex/0306021}{{\tt hep-ex/0306021}}.

\bibitem{Achard:2003ik}
{\bf L3} Collaboration, P.~Achard {\em et al.} {\em Phys. Lett.} {\bf B581}
  (2004) 19--30,
\href{http://www.arXiv.org/abs/hep-ex/0312026}{{\tt hep-ex/0312026}}.

\bibitem{Schael:2006ns}
{\bf ALEPH} Collaboration, S.~Schael {\em et al.}
\href{http://www.arXiv.org/abs/hep-ex/0604042}{{\tt hep-ex/0604042}}.

\bibitem{Siebel:2005uw}
M.~Siebel
\href{http://www.arXiv.org/abs/hep-ex/0505080}{{\tt hep-ex/0505080}}.

\bibitem{Golec-Biernat:1998js}
K.~Golec-Biernat and M.~Wusthoff {\em Phys. Rev.} {\bf D59} (1999) 014017,
\href{http://www.arXiv.org/abs/hep-ph/9807513}{{\tt hep-ph/9807513}}.

\bibitem{Golec-Biernat:1999qd}
K.~Golec-Biernat and M.~Wusthoff {\em Phys. Rev.} {\bf D60} (1999) 114023,
\href{http://arXiv.org/abs/hep-ph/9903358}{{\tt hep-ph/9903358}}.

\bibitem{Mueller:1993rr}
A.~H. Mueller {\em Nucl. Phys.} {\bf B415} (1994)
373--385.

\bibitem{Mueller:1994jq}
A.~H. Mueller and B.~Patel {\em Nucl. Phys.} {\bf B425} (1994) 471--488,
\href{http://www.arXiv.org/abs/hep-ph/9403256}{{\tt hep-ph/9403256}}.

\bibitem{Mueller:1994gb}
A.~H. Mueller {\em Nucl. Phys.} {\bf B437} (1995) 107--126,
\href{http://www.arXiv.org/abs/hep-ph/9408245}{{\tt hep-ph/9408245}}.

\bibitem{Salam:1996nb}
G.~P. Salam {\em Comput. Phys. Commun.} {\bf 105} (1997) 62--76,
\href{http://www.arXiv.org/abs/hep-ph/9601220}{{\tt hep-ph/9601220}}.

\bibitem{Salam:1995uy}
G.~P. Salam {\em Nucl. Phys.} {\bf B461} (1996) 512--538,
\href{http://www.arXiv.org/abs/hep-ph/9509353}{{\tt hep-ph/9509353}}.

\bibitem{Avsar:2004ms}
E.~Avsar
\href{http://www.arXiv.org/abs/hep-ph/0406150}{{\tt hep-ph/0406150}}.

\bibitem{Avsar:2005iz}
E.~Avsar, G.~Gustafson, and L.~Lönnblad {\em JHEP} {\bf 07} (2005) 062,
\href{http://www.arXiv.org/abs/hep-ph/0503181}{{\tt hep-ph/0503181}}.

\bibitem{Cohen-Tannoudji:1982pi}
G.~Cohen-Tannoudji, A.~Mantrach, H.~Navelet, and R.~Peschanski {\em Phys. Rev.}
  {\bf D28} (1983)
1628.

\bibitem{Mueller:1986ey}
A.~H. Mueller and H.~Navelet {\em Nucl. Phys.} {\bf B282} (1987)
727.

\bibitem{Gribov:1984tu}
L.~V. Gribov, E.~M. Levin, and M.~G. Ryskin {\em Phys. Rept.} {\bf 100} (1983)
1--150.

\bibitem{Capella:1992yb}
A.~Capella, U.~Sukhatme, C.-I. Tan, and J.~Tran Thanh~Van {\em Phys. Rept.}
  {\bf 236} (1994)
225--329.

\bibitem{Field:2005qt}
{\bf CDF} Collaboration, R.~Field {\em Acta Phys. Polon.} {\bf B36} (2005)
167--178.

\bibitem{Sjostrand:1987su}
T.~Sjostrand and M.~van Zijl {\em Phys. Rev.} {\bf D36} (1987)
2019.

\bibitem{Andersson:1983ia}
B.~Andersson, G.~Gustafson, G.~Ingelman, and T.~Sjostrand {\em Phys. Rept.}
  {\bf 97} (1983)
31.

\bibitem{Andersson:1991he}
B.~Andersson, G.~Gustafson, and C.~Sjogren {\em Nucl. Phys.} {\bf B380} (1992)
391--407.

\bibitem{Corcella:2000bw}
G.~Corcella {\em et al.} {\em JHEP} {\bf 01} (2001) 010,
\href{http://arXiv.org/abs/hep-ph/0011363}{{\tt hep-ph/0011363}}.

\bibitem{Sjostrand:2006za}
T.~Sjostrand, S.~Mrenna, and P.~Skands {\em JHEP} {\bf 05} (2006) 026,
\href{http://www.arXiv.org/abs/hep-ph/0603175}{{\tt hep-ph/0603175}}.

\bibitem{Lonnblad:1992tz}
L.~Lönnblad {\em Comput.~Phys.~Commun.} {\bf 71} (1992)
15--31.

\bibitem{Catani:1990yc}
S.~Catani, F.~Fiorani, and G.~Marchesini {\em Phys. Lett.} {\bf B234} (1990)
339.

\bibitem{Ciafaloni:1988ur}
M.~Ciafaloni {\em Nucl. Phys.} {\bf B296} (1988)
49.

\bibitem{Andersson:1995ju}
B.~Andersson, G.~Gustafson, and J.~Samuelsson {\em Nucl. Phys.} {\bf B467}
  (1996)
443--478.

\bibitem{Jung:2000hk}
H.~Jung and G.~P. Salam {\em Eur. Phys. J.} {\bf C19} (2001) 351--360,
\href{http://arXiv.org/abs/hep-ph/0012143}{{\tt hep-ph/0012143}}.

\bibitem{Jung:2001hx}
H.~Jung {\em Comput. Phys. Commun.} {\bf 143} (2002) 100--111,
\href{http://arXiv.org/abs/hep-ph/0109102}{{\tt hep-ph/0109102}}.

\bibitem{Kharraziha:1998dn}
H.~Kharraziha and L.~Lönnblad {\em JHEP} {\bf 03} (1998) 006,
\href{http://arXiv.org/abs/hep-ph/9709424}{{\tt hep-ph/9709424}}.

\bibitem{Kharraziha:ldcmc}
H.~Kharraziha and L.~Lönnblad {\em Comput. Phys. Commun.} {\bf 123} (1999) 153.

\bibitem{Kovchegov:1999yj}
Y.~V. Kovchegov {\em Phys. Rev.} {\bf D60} (1999) 034008,
\href{http://www.arXiv.org/abs/hep-ph/9901281}{{\tt hep-ph/9901281}}.

\bibitem{Balitsky:1995ub}
I.~Balitsky {\em Nucl. Phys.} {\bf B463} (1996) 99--160,
\href{http://www.arXiv.org/abs/hep-ph/9509348}{{\tt hep-ph/9509348}}.

\bibitem{Iancu:2002xk}
E.~Iancu, A.~Leonidov, and L.~McLerran
\href{http://www.arXiv.org/abs/hep-ph/0202270}{{\tt hep-ph/0202270}}.

\bibitem{Iancu:2003xm}
E.~Iancu and R.~Venugopalan
\href{http://www.arXiv.org/abs/hep-ph/0303204}{{\tt hep-ph/0303204}}.

\bibitem{Weigert:2005us}
H.~Weigert {\em Prog. Part. Nucl. Phys.} {\bf 55} (2005) 461--565,
\href{http://www.arXiv.org/abs/hep-ph/0501087}{{\tt hep-ph/0501087}}.

\bibitem{Jalilian-Marian:1997jx}
J.~Jalilian-Marian, A.~Kovner, A.~Leonidov, and H.~Weigert {\em Nucl. Phys.}
  {\bf B504} (1997) 415--431,
\href{http://www.arXiv.org/abs/hep-ph/9701284}{{\tt hep-ph/9701284}}.

\bibitem{Jalilian-Marian:1997gr}
J.~Jalilian-Marian, A.~Kovner, A.~Leonidov, and H.~Weigert {\em Phys. Rev.}
  {\bf D59} (1999) 014014,
\href{http://www.arXiv.org/abs/hep-ph/9706377}{{\tt hep-ph/9706377}}.

\bibitem{Iancu:2001ad}
E.~Iancu, A.~Leonidov, and L.~D. McLerran {\em Phys. Lett.} {\bf B510} (2001)
  133--144,
\href{http://www.arXiv.org/abs/hep-ph/0102009}{{\tt hep-ph/0102009}}.

\bibitem{Weigert:2000gi}
H.~Weigert {\em Nucl. Phys.} {\bf A703} (2002) 823--860,
\href{http://www.arXiv.org/abs/hep-ph/0004044}{{\tt hep-ph/0004044}}.

\bibitem{Mueller:1996te}
A.~H. Mueller and G.~P. Salam {\em Nucl. Phys.} {\bf B475} (1996) 293--320,
\href{http://www.arXiv.org/abs/hep-ph/9605302}{{\tt hep-ph/9605302}}.

\bibitem{Iancu:2004es}
E.~Iancu, A.~H. Mueller, and S.~Munier {\em Phys. Lett.} {\bf B606} (2005)
  342--350,
\href{http://www.arXiv.org/abs/hep-ph/0410018}{{\tt hep-ph/0410018}}.

\bibitem{Iancu:2003zr}
E.~Iancu and A.~H. Mueller {\em Nucl. Phys.} {\bf A730} (2004) 494--513,
\href{http://www.arXiv.org/abs/hep-ph/0309276}{{\tt hep-ph/0309276}}.

\bibitem{Iancu:2004iy}
E.~Iancu and D.~N. Triantafyllopoulos {\em Nucl. Phys.} {\bf A756} (2005)
  419--467,
\href{http://www.arXiv.org/abs/hep-ph/0411405}{{\tt hep-ph/0411405}}.

\bibitem{Mueller:2005ut}
A.~H. Mueller, A.~I. Shoshi, and S.~M.~H. Wong {\em Nucl. Phys.} {\bf B715}
  (2005) 440--460,
\href{http://www.arXiv.org/abs/hep-ph/0501088}{{\tt hep-ph/0501088}}.

\bibitem{Iancu:2005dx}
E.~Iancu, G.~Soyez, and D.~N. Triantafyllopoulos {\em Nucl. Phys.} {\bf A768}
  (2006) 194--221,
\href{http://www.arXiv.org/abs/hep-ph/0510094}{{\tt hep-ph/0510094}}.

\bibitem{Levin:2005au}
E.~Levin and M.~Lublinsky {\em Nucl. Phys.} {\bf A763} (2005) 172--196,
\href{http://www.arXiv.org/abs/hep-ph/0501173}{{\tt hep-ph/0501173}}.

\bibitem{Kozlov:2006cg}
M.~Kozlov, E.~Levin, and A.~Prygarin
\href{http://www.arXiv.org/abs/hep-ph/0606260}{{\tt hep-ph/0606260}}.

\bibitem{Kovner:2005aq}
A.~Kovner and M.~Lublinsky {\em Nucl. Phys.} {\bf A767} (2006) 171--188,
\href{http://www.arXiv.org/abs/hep-ph/0510047}{{\tt hep-ph/0510047}}.

\bibitem{Blaizot:2006wp}
J.~P. Blaizot, E.~Iancu, and D.~N. Triantafyllopoulos
\href{http://www.arXiv.org/abs/hep-ph/0606253}{{\tt hep-ph/0606253}}.

\bibitem{Praszalowicz:1997nf}
M.~Praszalowicz and A.~Rostworowski {\em Acta Phys. Polon.} {\bf B29} (1998)
  745--753,
\href{http://www.arXiv.org/abs/hep-ph/9712313}{{\tt hep-ph/9712313}}.

\bibitem{Lonnblad:1995wk}
L.~Lönnblad {\em Z.~Phys.} {\bf C65} (1995)
285--292.

\bibitem{Breitweg:2000yn}
{\bf ZEUS} Collaboration, J.~Breitweg {\em et al.} {\em Phys. Lett.} {\bf B487}
  (2000) 53--73,
\href{http://www.arXiv.org/abs/hep-ex/0005018}{{\tt hep-ex/0005018}}.

\bibitem{Adloff:2000qk}
{\bf H1} Collaboration, C.~Adloff {\em et al.} {\em Eur. Phys. J.} {\bf C21}
  (2001) 33--61,
\href{http://www.arXiv.org/abs/hep-ex/0012053}{{\tt hep-ex/0012053}}.

\bibitem{DIS04Petrukhin}
A.~Petrukhin, ``New Measurement of the Structure Function $F_2(x,Q^2)$ at low
  $Q^2$ with Initial State Radiation Data.'' Proceedings of DIS04,
  \v{S}trbsk\'{e} Pleso, Slovakia, 2004.

\bibitem{Sapeta:2005ba}
S.~Sapeta and K.~Golec-Biernat {\em Phys. Lett.} {\bf B613} (2005) 154--161,
\href{http://www.arXiv.org/abs/hep-ph/0502229}{{\tt hep-ph/0502229}}.

\end{thebibliography}\endgroup

\end{document}